\documentclass[amsmath,amssymb,reprint,nofootinbib,round]{revtex4-2}
\usepackage{mathptmx}
\usepackage[T1]{fontenc}
\usepackage[utf8]{inputenc}
\usepackage{xcolor}
\usepackage{varwidth}
\usepackage{amsmath}
\usepackage{graphicx}
\PassOptionsToPackage{normalem}{ulem}
\usepackage{ulem}
\usepackage[pdfusetitle,
 bookmarks=true,bookmarksnumbered=false,bookmarksopen=false,
 breaklinks=false,pdfborder={0 0 1},backref=false,colorlinks=false]
 {hyperref}

\makeatletter

\providecommand{\tabularnewline}{\\}
\newenvironment{cellvarwidth}[1][t]
    {\begin{varwidth}[#1]{\linewidth}}
    {\@finalstrut\@arstrutbox\end{varwidth}}
\providecolor{lyxadded}{rgb}{0,0,1}
\providecolor{lyxdeleted}{rgb}{1,0,0}
\DeclareRobustCommand{\mklyxadded}[1]{\bgroup\color{lyxadded}{}#1\egroup}
\DeclareRobustCommand{\mklyxdeleted}[1]{\bgroup\color{lyxdeleted}\mklyxsout{#1}\egroup}
\DeclareRobustCommand{\mklyxsout}[1]{\ifx\\#1\else\sout{#1}\fi}
\DeclareRobustCommand{\lyxadded}[4][]{\texorpdfstring{\mklyxadded{#4}}{#4}}

\makeatother

\begin{document}
\title{A practical guide to feedback control for Pound-Drever-Hall laser
linewidth narrowing}
\author{Wance Wang}
\affiliation{Department of Physics, University of Maryland, College Park, Maryland
20742, USA}
\author{Sarthak Subhankar}
\affiliation{Joint Quantum Institute, National Institute of Standards and Technology,
USA }
\affiliation{Department of Physics, University of Maryland, College Park, Maryland
20742, USA}
\author{Joseph W. Britton}
\affiliation{Department of Physics, University of Maryland, College Park, Maryland
20742, USA}
\affiliation{U.S. Army Combat Capabilities Development Command, Army Research Laboratory,
Adelphi, MD, 20783, USA}
\begin{abstract}
The Pound--Drever--Hall (PDH) technique for laser linewidth narrowing
is widely used by AMO experimentalists. However, achieving a high-performance
PDH locking requires substantial engineering experience, which is
scattered across literature and often lacks a cohesive control-theory
perspective. Excellent pedagogical papers exist on the theory of the
PDH error signal \citep{Black00ajop,Day92ijqe,Thorpe08oe,Reinhardt17oe},
but they rarely cover feedback control. General-purpose control theory
literature \citep{Bechhoefer05rmp,Ogata10PE} seldom discusses PDH
laser locking specifically. Although excellent PDH review articles
\citep{Hall99MP,Fox03NIST,Nagourney14OUP} provide thorough knowledge
and practice on both aspects, they are not novice-friendly. We extend
prior works \citep{Hall99MP,Fox03NIST,Nagourney14OUP} by addressing
component choice and loop tuning using modern tools like a vector
network analyzer. We organize multifaceted engineering considerations
systematically, grounded in feedback control principles. Our target
reader is researchers setting up a PDH laser lock for the first time;
we eschew advanced topics like minimizing residual amplitude modulation
(RAM). Our guidance is illustrated by step-by-step optimization of
the lock for a 1650 nm ECDL.
\end{abstract}
\maketitle

\section{Introduction}

The PDH optical frequency analyzer plays a central role in closed-loop
laser frequency stabilization and laser linewidth narrowing \citep{Drever83apb}.
The technique enables ultra-narrow line width lasers \citep{Young99prl,Ludlow07ol,Zhao10oc,Jiang11np,Kessler12np,Matei17prl},
record-setting optical atomic clocks \citep{Ludlow15rmp,Hinkley13s,Oelker19np,Steinel23prl,Kim23np,Aeppli24prl,Zhang24n}
and high-resolution spectroscopy \citep{Rafac00prl,Chwalla09Ths,Kleczewski12pra}.
Researchers at the forefront of laser stabilization can now build
ultra-narrow, high-stability cavities \citep{Ludlow07ol,Leibrandt11oe,Maddaloni13CP,Sanjuan19oe,Chen20col,Herbers22ol,Kelleher23oe}
and reduce effects like residual amplitude modulation (RAM) \citep{Zhang14ol,Shen15pra,Shi18apb,Bi19ao}
to obtain stabilization at the mHz level \citep{Matei17prl,Zhang17prl}.
It is our observation that many of the techniques used by expert practitioners
to obtain these impressive results are not well documented. We hope
this tutorial partially bridges this gap and broadens the accessibility
of laser frequency stabilization techniques.

\section{Just enough control theory}

We start with a brief introduction to feedback control theory and
omit details extraneous to the aim of first-order PDH optimization.
Excellent general-purpose control tutorials include those by Bechhoefer
\citep{Bechhoefer05rmp}, Ogata \citep{Ogata10PE}, and Doyle \citep{Doyle90MP}.
While preparing this manuscript, we noted another excellent summary
of both control theory and the PDH technique by Oswald \citep{Oswald22Ths}.

\subsection{The feedback loop}

\begin{figure}[th!]
\includegraphics[width=8.5cm]{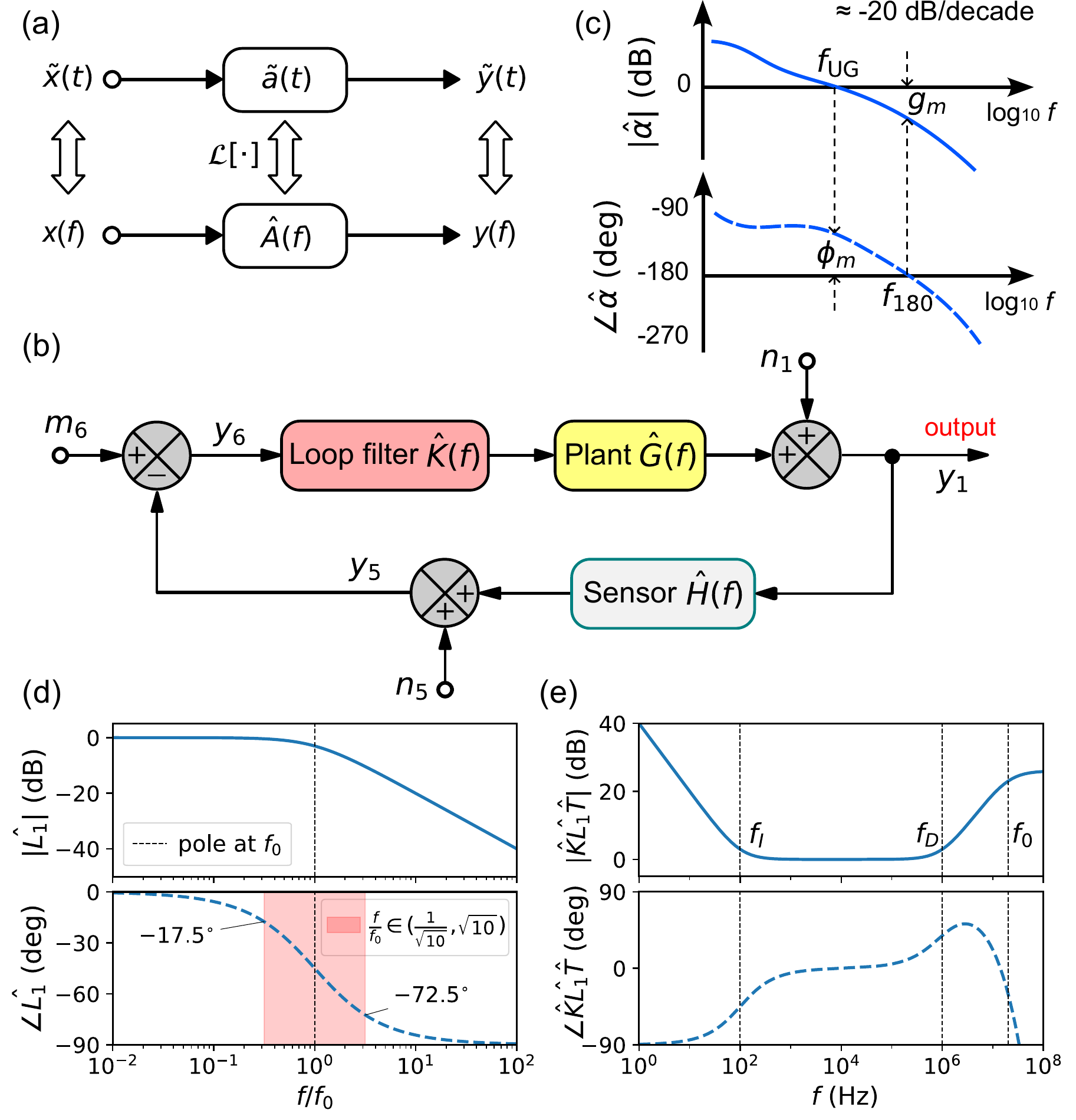}

\caption{\protect\label{fig:basics}(a) $\tilde{a}$ is a time-invariant (LTI)
system component that transforms signal $\tilde{x}$ to $\tilde{y}$.
The transfer function $\hat{A}$ encodes the frequency-domain behavior
of $\tilde{a}$. (b) The block diagram illustrates a feedback loop.
Components of the loop include a plant $\hat{G}$, sensor $\hat{H}$
and loop filter $\hat{K}$. Signals decorated by integers include
observable $y_{k}$, noise $n_{k}$ and modulation $m_{k}$ signals.
(c) A Bode plot for a sample open-loop transfer function $\hat{\alpha}$.
By convention the gain is plotted in dB ($20\log_{10}|\hat{\alpha}|$)
and the frequency is $\log_{10}f$. Dashed lines call out the \textit{unity-gain}
(UG) point $f_{\mathrm{UG}}$, \textit{phase margin} $\phi_{m}$,
\textit{phase crossover} point $f_{\mathrm{180}}$ and \textit{gain
margin} $g_{m}$. (d) The first-order low-pass filter $\hat{L}_{1}$
with corner frequency $f_{0}$ . It's properties include $3\text{ dB}$
attenuation and a $-45^{\circ}$ phase lag at $f_{0}$ . $60\%$ of
the phase change occurs in the one decade around $f_{0}$ (red). (e)
A sample plot of a PID loop filter $\hat{K}\hat{L}_{1}\hat{T}$ transfer
function where $\hat{K}$ is given by Eq.$\,$\ref{eq:PID}, $K_{P}=1$,
$f_{I}=100\text{ Hz}$, and $f_{D}=1\text{ MHz}$. In the high-frequency
limit all devices are ultimately bandwidth-limited and exhibit low-pass
behavior, corner frequency $f_{0}=20\text{ MHz}$ for $\hat{L}_{1}$
and delay $\tau_{l}=10\text{ ns}$ for $\hat{T}$.}
\end{figure}

Our system consists of a collection of components and signals. All
components are linear and time-invariant (LTI). Component $\tilde{a}(t)$
transforms an input signal $\tilde{x}(t)$ to an output signal $\tilde{y}(t)$.
Their relation is simple in Laplace domain
\begin{equation}
\hat{A}=y/x.
\end{equation}
$\hat{A}(f)=\mathcal{L}\left[\tilde{a}(t)\right]$ is called the transfer
function of the component (Fig.$\:$\ref{fig:basics}(a)) and we set
the complex variable $s=j2\pi f$. We use the $\sim$ symbol for time-domain
signals and $\text{\textasciicircum}$ symbol for transfer functions.
To reduce clutter we often omit the $\mbox{\textasciicircum}$ for
Laplace-domain signals $x(f)=\mathcal{L}\left[\tilde{x}(t)\right]$
and the explicit time or frequency dependence. The magnitude and phase
of $\hat{A}$ are denoted by $|\hat{A}|$ and $\angle\hat{A}$, respectively.
As $f$ increases, the change in $\angle\hat{A}$ is called the \textit{phase
shift}. When $\angle\hat{A}>0$ it is \textit{phase lead} and when
$\angle\hat{A}<0$ it is \textit{phase lag}. A vector network analyzer
(VNA) is a tool used in system discovery: by driving $x$ and observing
$y$ we learn about $\hat{A}$. It is common to visualize transfer
functions with a Bode plot (Fig.$\:$\ref{fig:basics}(c)).

To implement a feedback control system, components are connected as
a loop. The loop signals $\tilde{y}_{k}(t)$ can represent a range
of physical quantities including electrical (in volts) and optical
frequency (in Hz). In frequency domain, $y_{k}=\mathcal{L}\left[\tilde{y}_{k}(t)\right]$.
A block diagram (Fig.$\:$\ref{fig:basics}(b)) is a pictorial representation
of transfer functions and signals $y_{k}$, where $k$ labels the
signal position in the block diagram. Common loop components are the
plant $\hat{G}$, sensor $\hat{H}$ and loop filter $\hat{K}$. We
call signal $y_{1}$ the \textit{output signal} and $y_{5}$ the \textit{error
signal}. The signal $m_{6}$ is an external input. We represent noise
in the plant and sensor as signals $n_{1}$ and $n_{5}$ (Fig.$\:$\ref{fig:basics}(d)).

The \textit{open-loop transfer function }\textit{\emph{is}} \textit{\emph{the
product of all loop components }}\citep{Nagourney14OUP}
\begin{equation}
\hat{\alpha}=\hat{K}\hat{G}\hat{H}\label{eq:open}
\end{equation}
\textit{The closed-loop transfer function} is the ratio of error signal
$y_{5}$ to $m_{6}$ 
\begin{equation}
\frac{y_{5}}{m_{6}}=\frac{\hat{K}\hat{G}\hat{H}}{1+\hat{K}\hat{G}\hat{H}}=\frac{\hat{\alpha}}{1+\hat{\alpha}}
\end{equation}
If we add noises $n_{1}$ and $n_{5}$, the output becomes
\begin{equation}
y_{1}=\frac{\hat{\alpha}}{1+\hat{\alpha}}\left(\frac{m_{6}}{\hat{H}}+\frac{n_{1}}{\hat{\alpha}}-\frac{n_{5}}{\hat{H}}\right)\label{eq:tension}
\end{equation}
The power spectral density (PSD) is 
\begin{equation}
S_{y1}=\frac{1}{\left|1+\hat{\alpha}\right|^{2}}S_{n1}+\frac{\left|\hat{\alpha}\right|^{2}}{\left|1+\hat{\alpha}\right|^{2}}\frac{1}{\left|\hat{H}\right|^{2}}S_{n5}\label{eq:PSD_loop}
\end{equation}
where $S_{n1}$ and $S_{n5}$ are the noise PSD.

Ideal \textit{negative feedback} is characterized by a loop where
$|\hat{\alpha}|\rightarrow\infty$ while $\angle\hat{\alpha}=0^{\circ}$
at all frequencies of interest. In this case we have strong suppression
of $n_{1}$: $y_{1}=(m_{6}-n_{5})/\hat{H}$. However, there is no
suppression of $n_{5}$. This tension between high gain reducing laser
noise but increasing susceptibility to measurement noise will be a
recurring theme.

If at any frequency $|\hat{\alpha}(f)|>1$ and $\angle\hat{\alpha}(f)<-150^{\circ}$,
the loop amplifies noise and may be unstable. Given the central role
of $\hat{\alpha}$ in loop performance, the control community has
developed terminologies to describe loop stability \citep{Ogata10PE}
(see Fig.$\:$\ref{fig:basics}(c)). The \textit{unity-gain} (UG)
point $f_{\mathrm{UG}}$ is the frequency where $|\hat{\alpha}(f_{\mathrm{UG}})|=0\,\mathrm{dB}$\footnote{$f_{\mathrm{UG}}$ is also called ``loop bandwidth'' \citep{Seiler21Doc}.}
and the \textit{phase margin} is $\phi_{m}=180^{\circ}+\angle\hat{\alpha}(f_{\mathrm{UG}})$.
At $f_{\mathrm{UG}}$ with an empirical $\phi_{m}=60^{\circ}$, the
factor $1/\left|1+\hat{\alpha}\right|^{2}=1$ in Eq.$\:$\ref{eq:PSD_loop},
so the loop has no suppression on the noise spectrum $S_{n1}$ when
$f>f_{\mathrm{UG}}$. The \textit{phase crossover} point $f_{180}$
is the frequency where $\angle\hat{\alpha}(f_{180})=-180^{\circ}$
and \textit{gain margin} is $g_{m}=1/|\hat{\alpha}(f_{\mathrm{180}})|$.
A consequence of negative feedback is increased sensitivity to noise
called the \emph{servo bump} \citep{Li22pra}, which commonly occurs
over the range $f_{\mathrm{UG}}<f<f_{180}$. We define $f_{\mathrm{bump}}$
to be the frequency where servo bump noise spectral density is a maximum
\footnote{If the $S_{n5}$ term in Eq.$\:$\ref{eq:PSD_loop} is ignored, the
position of $f_{\mathrm{bump}}$ is where $\left|1+\hat{\alpha}\right|^{2}$
is minimum so $S_{y1}$ is maximized.}.

\subsection{Loop components\protect\label{subsec:Loop-Components}}

Now we introduce some components found in PDH laser locks (Fig.$\:$\ref{fig:basics}(b)).
\begin{description}
\item [{plant}] The plant is a laser. We define the laser optical frequency
(in Hz) to be $\tilde{y}_{1}+\nu_{c}$, where $\nu_{c}$ is an optical
cavity resonance (at an optical frequency) and $\tilde{y_{1}}$ is
the laser-cavity detuning (at a RF frequency). The signal $y_{1}=\mathcal{L}\left[\tilde{y}_{1}\right]$
is the Laplace transform of the laser detuning and $n_{1}$ is its
frequency noise. 
\item [{sensor}] We need a sensor that transforms frequency signal $\tilde{y}_{1}$
into voltage signal $\tilde{y}_{5}$. Here we focus on the cavity-based
PDH optical frequency discriminator (the ``PDH discriminator'')
\citep{Hall99MP,Black00ajop}. This sensor measures an optical frequency
relative to an internal, non-adjustable frequency $\nu_{c}$ (the
nearest cavity transmission feature). Its output is a voltage $\tilde{y}_{5}\propto\tilde{y}_{1}$
and its transfer function is $\hat{H}_{\nu_{c}}$. Another property
of $\hat{H}_{\nu_{c}}$ is that its output is linear only for $\tilde{y}_{1}\in(-\delta\nu_{c}/2,\delta\nu_{c}/2)$
, where $\delta\nu_{c}$ is the cavity optical linewidth (Fig.$\:$\ref{fig:error-drift}(b)).

Note that outside this bandwidth the LTI assumption fails. Measuring
$\hat{H}_{\nu_{c}}$ directly is non-trivial as many lasers require
active feedback to remain within $\nu_{c}\pm\delta\nu_{c}/2$ --
we will return to this topic later.
\item [{loop filter}] The transfer function of a loop filter $\hat{K}$
is tailored to advance feedback design goals. The proportional--integral--derivative
(PID) loop filter is widely-used 
\begin{equation}
\hat{K}=K_{P}\left(1-j\frac{f_{I}}{f}+j\frac{f}{f_{D}}\right)\label{eq:PID}
\end{equation}
where $K_{P}$ is proportional gain, $f_{I}$, $f_{D}$ are the P-I
and P-D corner frequencies, respectively (Fig.$\:$\ref{fig:basics}(d)).
The second term is an integrator (I) with gain slope $-20\,\mathrm{dB}$
per decade that adds $-90^{\circ}$ phase lag when $f\ll f_{I}$.
The third term is a differentiator (D) with gain slope $+20\,\mathrm{dB}$
per decade that adds $+90^{\circ}$ phase lead when $f\gg f_{D}$.
We adjust the phase lag of $\hat{K}$ by tuning $f_{\text{I}}$ and
$f_{D}$. The UG point can be extended by using the up to $+90^{\circ}$
phase lead available from the $D$ term. \footnote{Real-world PID devices often fall short of the full $+90^{\circ}$
phase lead available from an ideal $D$ term. See the plot Fig.$\:$\ref{fig:basics}(e).} The control available from PID is usually enough but some lasers
may benefit from more sophisticated feedback \citep{Ryou17rosi}.
\item [{delay}] An implicit element of any feedback loop is propagation
delay $\tau_{l}$. The transfer function is $\hat{T}=\exp\left(-j2\pi f\tau_{l}\right)$
which has magnitude $1$ and a phase that increases linearly in $f$
without bound. The phase delay $2\pi f\tau_{l}$ is fundamental and
can't be corrected by other loop components. At high frequency, the
phase lag due to $\hat{T}$ can limit $f_{\mathrm{UG}}$ and feedback
performance. For example, the phase lag is $18^{\circ}$ at $f=1\,\mathrm{MHz}$
if the loop length is $10\,\mathrm{m}$ in coaxial cables. The transfer
function $\hat{T}$ does not explicitly appear in block diagrams as
it is distributed across all components.\\
\end{description}
Simple models for these components include the following.
\begin{description}
\item [{first-order low-pass }] The low-pass filter is a common loop element
that results in a phase shift. The transfer function for a first-order
low-pass is (Fig.$\:$\ref{fig:basics}(d)), 
\begin{equation}
\hat{L}_{1}=\frac{1}{1+jf/f_{0}}
\end{equation}
where $f_{0}$ is the corner frequency. The attenuation of $\hat{L}_{1}$
at $f_{0}$ is $-3\mbox{ dB}$ and for $f>f_{0}$ the slope is $-20\mbox{ dB}$/decade.
The phase shift is $\angle\hat{L}_{1}=-\arctan(f/f_{0})$. Notice
that $\angle L_{1}(f_{0})$ is $-45^{\circ}$ and it has the steepest
change in the decade $(1/\sqrt{10},\ensuremath{\sqrt{10}})f_{0}$.
The electronic RC circuit is an example of a first-order low-pass
where $f_{0}=1/(2\pi RC)$. 
\item [{high-order low-pass}] Later we will encounter a need for higher-order
low-pass filters $\hat{L}_{n}$ with corner $f_{0}$, and where $\lim_{f\rightarrow\infty}\angle\hat{L}_{n}=-90^{\circ}\times n$
\citep{Nagourney14OUP}. The common $n$-th order Butterworth low-pass
$\widehat{LB}_{n}$ has slope $-20\,\mathrm{dB}\times n$ per decade
for $f>f_{0}$. At intermediate frequencies $\angle\widehat{LB}_{n}$
can be computed numerically \citep{scipy} or approximated by $\angle\widehat{LB}_{n}\approx-n\arctan(f/f_{0})$.
Strong attenuation comes at the cost of large phase lag -- the right
filter is context dependent. All real-world plants $\hat{G}$ exhibit
low-pass behavior at high frequency and can be approximated by some
low-pass $\hat{L}_{n}$. 
\end{description}

\subsection{Feedback design goals \protect\label{subsec:Goals}}

Here we provide advice for choosing feedback parameters in simple
closed-loop systems. Our high-level design goals are the following:
\begin{enumerate}
\item maximize $f_{\mathrm{UG}}$
\item maintain $30^{\circ}<\phi_{m}<60^{\circ}$
\item for $f<f_{\mathrm{UG}}$, maintain $\angle\hat{\alpha}>-120^{\circ}$
\end{enumerate}
Naturally, for $f<f_{\text{UG}}$ we also want to maximize $|\hat{\alpha}|$.
However, this aim is constrained by a gain-phase relationship first
considered by Bode \citep{Bode40bstj,Bechhoefer05rmp}, a special
case of the Kramers-Kronig relation \citep{Bechhoefer11ajp}.
\begin{quote}
For a transfer function $\hat{A}$\footnote{$\hat{A}$ has to be minimum-phase \citep{Bechhoefer11ajp}},
phase $\angle\hat{A}$ is uniquely related to gain $|\hat{A}|$. When
the slope of $20\log_{10}|\hat{A}(f)|$ is $-n$, the phase $\angle\hat{A}(f)\approx-n\times90^{\circ}$.
\end{quote}
This relation dictates that the open-loop gain $|\hat{\alpha}|$ should
decrease by at most $-20\text{ dB/decade}$ to avoid $\angle\hat{\alpha}<-150^{\circ}$;
note that our design goals 2 and 3 are even more strict constraints
in the interest of maintaining loop stability. This rationalizes our
first design goal: because the slope of $|\hat{\alpha}|$ is constrained,
the only way to obtain large $|\hat{\alpha}|$ is to maximize $f_{\text{UG}}$.
A corollary is that the primary way to add phase lead near $f_{\text{UG}}$
is to increase $|\hat{\alpha}|$. Because of the phase lag in Fig.$\:$\ref{fig:basics}(d),
in many cases the highest achievable $f_{\mathrm{UG}}$ is one decade
lower than $f_{0}$ for the dominant low-pass in the loop. Note that
the feedback can be made more robust to long-term drift in $|\hat{\alpha}|$
by minimizing the slope $d\angle\hat{\alpha}/df$ near $f_{\mathrm{UG}}$.

Two key trade-offs should be considered when tuning the loop phase
margin. First, an increase in $|\hat{\alpha}|$ is accompanied by
a reduction in $\phi_{m}$. That is, better suppression of noise $n_{1}$
in one frequency band comes at the cost of loop instability and stronger
servo bump in another. Second, small $\phi_{m}$ provides fast transient
response but suffers from increased ringing and overshoot. In practice,
empirical range $30^{\circ}<\phi_{m}<60^{\circ}$ is recommended for
satisfactory performance \citep{Ogata10PE}.

The last design goal ensures stability for all $f<f_{\mathrm{UG}}$.
The Ziegler-Nichols method \citep{Ogata10PE} is an algorithm for
tuning PID parameters that respects these design goals but is typically
not optimal -- we return to this topic in Sec.\ref{sec:system-optimization-routine}.

\section{PDH lock component choices}

\label{sec:PDH-Lock-Component-Choices}

In this section we discuss a laser frequency stabilization system
containing a PDH optical frequency discriminator (OFD) $\hat{H}_{\nu_{c}}$
and provide advice on component choices. The system consists of loop
signals $y_{k}$ and the following components (Fig.$\:$\ref{fig:setups}(a)):
a laser to be stabilized, an electro-optic modulator (EOM), an optical
reference cavity, a photodetector (PD), RF demodulation electronics
and a loop filter. For details on the PDH frequency discriminator
itself, the reader is referred to Ref.$\:$\citep{Black00ajop} and
Sec.$\:$\ref{subsec:PDH-discriminator}.

\subsection{Laser}

To be concrete we consider a generic external-cavity diode laser (ECDL).
Output signal $y_{1}$ is the laser frequency which is modulated by
two actuators: laser diode current (the \textit{fast branch}) and
grating diffraction angle modulated by PZT (the \textit{slow branch}).
Each feedback path can be modeled by its own open-loop transfer function
($\hat{G}_{\mathrm{fast}}$ and $\hat{G}_{\mathrm{slow}}$) with laser-limited
low-pass corner frequencies ($f_{0\mathrm{,fast}}$ and $f_{0\mathrm{,slow}}$).
The high-pass corner frequency of the \textit{fast branch} is $f_{\mathrm{HP,fast}}$,
typically set by the laser diode head board \citep{Preuschoff22rosi}.

We assume a simple model for free-running laser frequency noise \citep{Domenico10ao,Bandel16oe,Nagourney14OUP}.
Let $S_{n1}(f)$ be the frequency noise PSD ($\mbox{Hz}^{2}/\mbox{Hz}$)
\citep{Nagourney14OUP}. This quantity can be calculated from the
auto-correlation $R_{\phi}(t)$ of the laser phase over time $\phi(t)$.
\begin{equation}
R_{\phi}(\tau)=\lim_{T\rightarrow\infty}\frac{1}{T}\int_{T/2}^{T/2}\phi(t+\tau)\phi(t)dt
\end{equation}
\begin{equation}
S_{n1}(f)=\frac{1}{2\pi}f^{2}\int_{\infty}^{\infty}R_{\phi}(\tau)e^{i2\pi f\tau}d\tau
\end{equation}
Free-running laser noise can be parameterized as 
\begin{equation}
S_{n1}(f)=h_{-1}f^{-1}+h_{0}f^{0}\label{eq:laser-power-law}
\end{equation}
where $h_{-1}$ is the 1-Hz-frequency intercept for $1/f$-noise and
$h_{0}$ is the 1-Hz-frequency intercept for white frequency noise.
The instantaneous laser linewidth is dominated by the white noise
and has a Lorentzian line shape with $\pi h_{0}$ full-width at half-max.
The $\beta\text{-separation}$ method \citep{Domenico10ao} is a technique
for estimating the linewidth for noise spectra that do not necessarily
follow Eq.$\:$\ref{eq:laser-power-law}. We note that a single linewidth
number can be quite misleading \citep{Bandel16oe,Fox03NIST,Bai21fp,Chen24s};
we recommend obtaining the spectrum $S_{n1}(f)$ from laser manufacturers
prior to purchasing a laser and PDH stabilization components (especially
the reference cavity). 

Although we focus on ECDL in this work, the description here captures
features that can be similar in other types of lasers. For background
on other laser types, readers may refer to Ref.$\:$\citep{Maddaloni13CP,Nagourney14OUP}.

We recommend the following properties for an ECDL to be easily PDH-stabilized:
\begin{itemize}
\item $f_{0\mathrm{,fast}}$ is at least one decade higher than the noise
that one wants to suppress
\item $f_{\mathrm{HP,fast}}$ should be lower than $f_{0\mathrm{,slow}}/10$
\end{itemize}
The first is motivated by the observation that $f_{\mathrm{UG}}\sim f_{0\mathrm{,fast}}/10$.
Further, since $f_{\mathrm{UG}}$ is often limited to $\sim2\,\mathrm{MHz}$
by loop propagation delay, we recommend $f_{0\mathrm{,fast}}>10\text{ MHz}$.
However, too strong $S_{n1}(f)$ may still preclude stable laser locking
even though $f_{\text{UG}}$ is faster than the noise, see our analysis
in Ref.$\:$\citep{wang2024pdh-lockable}.

Distinct fast- and slow-modulation inputs are present in most laser
systems. For many laser types the slow feedback path $\hat{G}_{\text{slow}}$
has many resonances as it is effected by PZT actuation of mechanical
components like mirrors or gratings \citep{Briles10oe,Nakamura20oe}.
The second is motivated by the observation that feedback via $\hat{G}_{\text{fast}}$
needs to suppress the slow branch servo bump so we require $f_{\mathrm{HP,fast}}<f_{0\mathrm{,slow}}/10$.
Note that advanced techniques that can eliminate the servo bump altogether
\citep{Li22pra}.

\subsection{PDH frequency discriminator}

Generation of the PDH error signal $\tilde{y}_{5}$ relies on several
components which can be independently optimized.

\subsubsection{Modulation and demodulation \protect\label{sec:Mod} }

An EOM driven at RF frequency $\Omega$ is used to add side bands
to the laser. An RF mixer demodulates the PD output using $\Omega$
applied to the LO port of a mixer. There is a low-pass filter downstream
of the mixer with corner frequency $f_{M}$ that outputs signal $\tilde{y}_{5}$.
The demodulation transfer function (mixer+low-pass) is $\hat{D}$.

Common choices for $\Omega/2\pi$ are $10$ to $20\mbox{ MHz}$ \citep{Alnis08pra,Sherstov10pra,Zhao10oc,Zhang14ol}.
Resonant EOMs operate at a single manufacturer-defined frequency,
while fiber-EOMs have roughly uniform response from $1$ MHz to beyond
$10$ GHz, they have lower power handling. For laser light at frequency
$\tilde{y}_{1}$ the EOM output is $\exp(j2\pi\tilde{y}_{1}t+j\beta\sin(\Omega t))$
which can be expanded as

\begin{equation}
e^{j2\pi\tilde{y}_{1}t}\left(J_{0}(\beta)+\sum_{k=1}^{\infty}J_{k}(\beta)e^{jk\Omega t}+\sum_{k=1}^{\infty}(-1)^{k}J_{k}(\beta)e^{-jk\Omega t}\right)
\end{equation}
where $\beta$ is the modulation depth and $J_{k}$ is $k$-th order
Bessel function. Mixing of the side bands $J_{0}$ and $J_{\pm1}$
reflected by the optical cavity gives rise to signal $\tilde{y}_{5}$.
If $\beta=1.082$ the linear region slope $k_{e}$ (discussed later
in Eq.$\:$\ref{eq:ke}) is maximum. A DC term in $\tilde{y}_{4}$
is blocked by the RF port of the mixer but $2\Omega$ terms are demodulated
to frequency $\Omega$ in $\tilde{y}_{5}$. The desired signal $\tilde{y}_{5}$
is proportional to $q=4J_{0}(\beta)J_{1}(\beta)$ while unwanted modulation
at $\Omega$ is proportional to $p=2J_{1}(\beta)^{2}+4J_{0}(\beta)J_{2}(\beta)$
\citep{Black00ajop}. A low-pass filter after the mixer blocks the
unwanted modulation at the cost of a phase shift that limits $f_{\text{UG}}$\footnote{Some loop filters provide sufficient attenuation at $\Omega$ and
don't require an external low-pass.}. To reach a signal-to-noise ratio (SNR) of $10^{3}$, attenuation
$-20\log_{10}(10^{3}/\frac{q}{p})\approx-55\mbox{ dB}$ is needed
at $\Omega$.

Suppose that $\Omega/2\pi=20\mbox{ MHz}$ and $f_{\mathrm{UG}}=1\mbox{ MHz}$.
An $n$-th order Butterworth filter provides $-55\mbox{ dB}$ attenuation
when $f_{M}=(\Omega/2\pi)10^{-\frac{55}{20n}}$. Numerical results
show that the attenuation target is met when $n=8$ and $f_{M}=9\mbox{ MHz}$;
the phase lag is $\angle\widehat{LB}_{n}(f_{\mathrm{UG}})=-33^{\circ}$.
To reduce the phase lag one has to either increase modulation $\Omega$
or sacrifice SNR. Alternative solutions include using a high-Q notch
filter at $\Omega$ \citep{Fox03NIST} or relying on the loop filter's
inherent low-pass characteristics while removing additional filters.

\subsubsection{Photodetector}

The light detector consists of a photodetector and a transimpedance
amplifier. We call this composite device a PD and call its transfer
function $\hat{P}$. The relevant properties are its responsivity
$R$ (in units of $\mathrm{A/W}$), transimpedance gain $g_{\mathrm{tr}}$
(in units of $\mathrm{V/A}$), noise equivalent power (NEP, in units
of $\mathrm{W/\sqrt{Hz}}$) and its frequency response, modeled as
an $n$-th order low-pass $\hat{L}_{\mathrm{PD}}$ with corner frequency
$f_{\mathrm{PD}}$. For incident optical power $P_{\mathrm{PD}}$,
the output voltage amplitude at DC is $P_{\mathrm{PD}}Rg_{\mathrm{tr}}$.

Noise in the PD can be consequential to overall PDH performance. Contributions
include detector electronic noise (including electronic shot noise)
characterized by NEP \citep{Mackowiak15Doc} and photon shot noise
\citep{Day92ijqe,Black00ajop}. The NEP is defined as the incident
power that gives an SNR of one in a $1\mbox{ Hz}$ bandwidth. For
photocurrent, the RMS electronic noise in bandwidth $\delta$ is $I_{en}=\mbox{NEP}\times R\sqrt{\delta}$
and the noise by photon shot noise at power $P_{\mathrm{PD}}$ is
$I_{sn}=\sqrt{4eRJ_{1}(\beta)^{2}P_{\mathrm{PD}}}\sqrt{\delta}$.
The optical power when $I_{en}=I_{sn}$ is $P_{\mathrm{PD}}^{(eq)}=\mbox{NEP}^{2}R/(4eJ_{1}(\beta)^{2})$.
The SNR at $P_{\mathrm{PD}}^{(eq)}$ is then
\begin{equation}
\frac{P_{\mathrm{PD}}^{(eq)}R}{I_{en}+I_{sn}}=\frac{\mbox{NEP}}{8eJ_{1}(\beta)^{2}\sqrt{\delta}}R
\end{equation}
The threshold $P_{\mathrm{PD}}>P_{\mathrm{PD}}^{(eq)}$ is notable
since above it electronic noise no longer dominates. In the context
of PDH, the bandwidth that contributes to noise in $y_{5}$ is $(\Omega/2\pi-f_{M},\Omega/2\pi+f_{M})$,
where $f_{M}$ is the corner frequency of the low-pass following the
mixer. Now we can put some numbers to it: suppose NEP $=10\mbox{ pW}/\sqrt{\mbox{Hz}}$,
$R=1\,\mathrm{A/W}$ and $f_{M}=9\mbox{ MHz}$. Then, $P_{\mathrm{PD}}^{(eq)}=720\text{ \ensuremath{\mu\text{W}}}$
and the $\mbox{SNR}$ is $8480$.

In Appendix$\,$\ref{appendix:lock-in-detection}, we show that $\hat{P}$
depends on the value of $\Omega$ due to the mixer:
\begin{equation}
\hat{P}=e^{\frac{j}{2}\left[\angle\hat{L}_{\mathrm{PD}}(\Omega+f)-\angle\hat{L}_{\mathrm{PD}}(\Omega-f)\right]}\label{eq:PD-phase}
\end{equation}
We see wide variation in the phase response of commercial PD packages
that can be approximated by $\angle\hat{L}_{\mathrm{PD}}(f)\approx-n\arctan(f/f_{\mathrm{PD}})$,
where $n$ is the number of poles. We find $n=3$ for several PDs
in our lab. Given $\Omega/2\pi=20\,\mathrm{MHz}$, $f_{\mathrm{PD}}=10\Omega/2\pi$
and $f_{\mathrm{UG}}=1\mbox{ MHz}$, we get $\angle\hat{P}(f_{\mathrm{UG}})=-0.9^{\circ}$.
An especially sub-optimal choice is $\Omega/2\pi=5\,\mathrm{MHz}$
and $f_{\mathrm{PD}}=5\,\mathrm{MHz}$ where $\angle\hat{P}(f_{\mathrm{UG}})=-17^{\circ}$.

Given what we've seen so far you might think $\Omega/2\pi>20\mbox{ MHz}$
is an improvement, however it's not so simple. Nonidealities that
emerge for such frequencies include the following. 1) The phase lag
$\angle\hat{P}$ becomes larger. 2) The electronics for offset-locking
becomes marginally harder \citep{Thorpe08oe,Bai17jo,Rabga23oe,Tu24arX,Hildebrand24arX}.
3) It is increasingly hard to avoid overlap of EOM sidebands with
higher-order transverse cavity modes \citep{Anderson84ao}.

\subsubsection{Reference cavity}

An ultrastable Fabry-Perot optical cavity is commonly used in PDH-based
frequency discriminators \citep{Maddaloni13CP,Nagourney14OUP}. Salient
cavity properties are its transmission frequency $\nu_{c}$, the resonance
full width at half max linewidth $\delta\nu_{c}$ and free spectral
range (FSR) $\nu_{\mathrm{FSR}}$. The cavity finesse is $\mathcal{F}=\nu_{\mathrm{FSR}}/\delta\nu_{c}$
and the resonator quality factor is $Q=\frac{c/\lambda}{\delta\nu_{\text{c}}}$.
The transfer function for the light reflected by the cavity has a
low-pass response 
\begin{equation}
\hat{C}=1/(1+j2f/\delta\nu_{c})\label{eq:cavity-transfer-function}
\end{equation}
\citep{Day92ijqe,Nagourney14OUP,Reinhardt17oe}, where $\delta\nu_{c}/2$
is the corner frequency. Appendix$\:$\ref{appendix:ringdown} shows
how to measure $\delta\nu_{c}$ by cavity ring-down. Guidance on laser
beam alignment to the cavity is in \citep{Anderson84ao,Boyd24ajop}.\lyxadded{JWB}{Tue Dec 10 20:00:04 2024}{
}

\subsubsection{Discriminator \protect\label{subsec:PDH-discriminator}}

Now we combine our knowledge of individual components to better understand
properties of the PDH frequency discriminator transfer function $\hat{H}_{\nu_{c}}$.
The subscript $\nu_{c}$ reminds us that the reference frequency is
defined by the cavity.

\begin{figure}
\includegraphics[width=7.5cm]{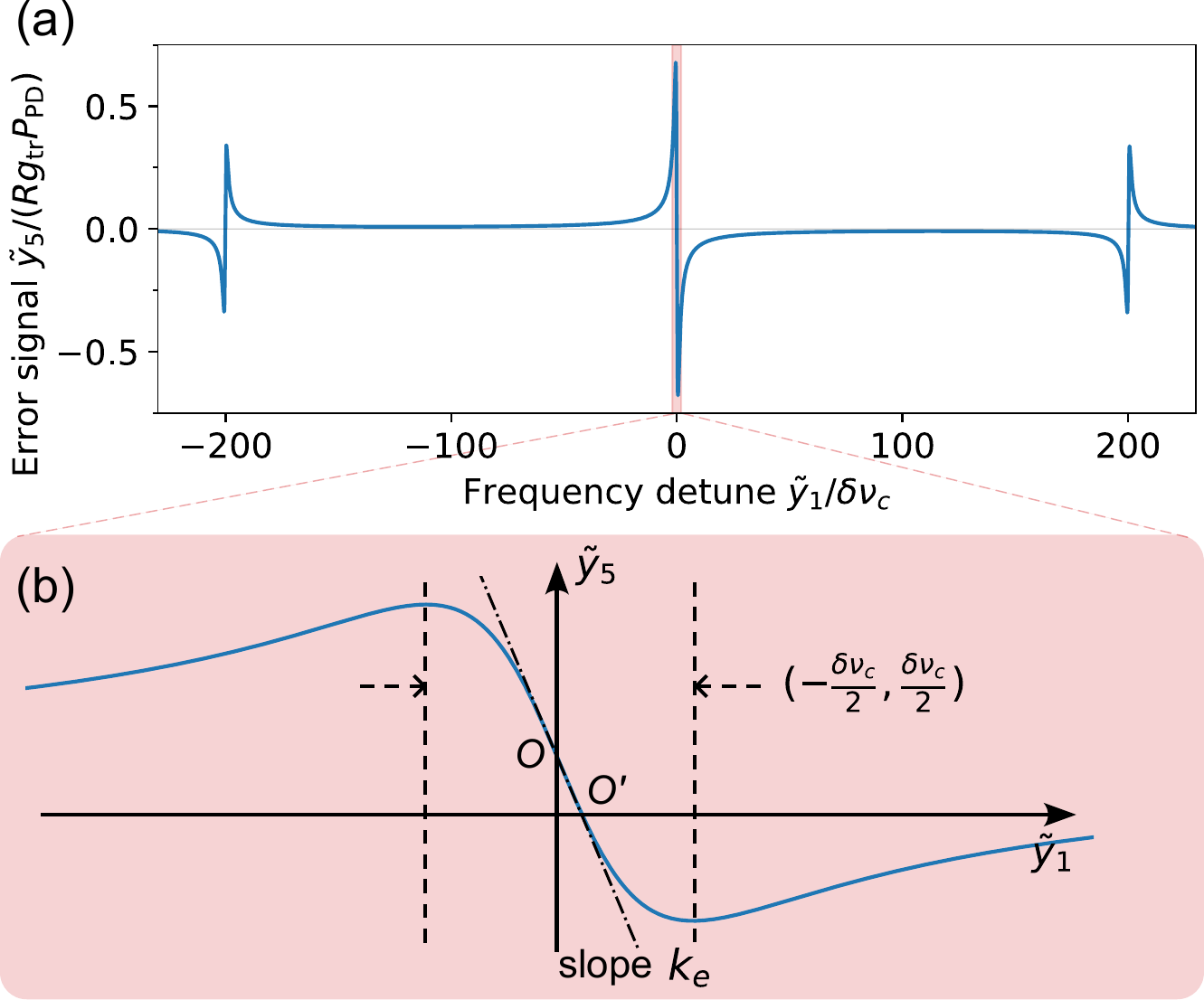}

\caption{\protect\label{fig:error-drift}(a) A sample PDH error signal for
$\delta\nu_{c}=100\text{ kHz}$, $\Omega/2\pi=20\text{ MHz}$ and
$\beta=1.082$. $\tilde{y}_{1}$ and $\tilde{y}_{5}$ are instantaneous
laser frequency detuning and error signal in time domain. (b) Zoom-in
of the red-shaded area in (a). Signal $\tilde{y}_{5}$ is linear only
over a narrow region $\tilde{y}_{1}\in(-\delta\nu_{c}/2,\delta\nu_{c}/2)$.
Ideally the signal is symmetric about the zero crossing point $O$.
Here we show an offset error: $\tilde{y}_{5}=0$ is at $O^{\prime}$.}
\end{figure}

A sample PDH error signal is plotted in Fig.$\:$\ref{fig:error-drift}(a).
The slope $k_{e}$ of the discriminator response is \citep{Black00ajop}
\begin{equation}
k_{e}:=\left.\frac{d\tilde{y}_{5}}{d\tilde{y}_{1}}\right|_{\tilde{y}_{1}=0}=8J_{0}(\beta)J_{1}(\beta)P_{\mathrm{PD}}Rg_{\mathrm{tr}}/\delta\nu_{c}\label{eq:ke}
\end{equation}
with units of $\mbox{V/Hz}$. Note that $\tilde{y}_{1}$ and $\tilde{y}_{5}$
are time-domain signals. The overall PDH discriminator transfer function
is given by (see Fig.$\:$\ref{fig:setups}(b)) 
\begin{equation}
\hat{H}_{\nu_{c}}:=\frac{y_{5}}{y_{1}}\approx\hat{D}\hat{P}k_{e}\hat{C}\label{eq:PDH-transfer}
\end{equation}
and includes terms for demodulation $\hat{D}$, photodetector $\hat{P}$
and cavity $\hat{C}$.

Following are several design goals to optimize $\hat{H}_{\nu_{c}}$.
\begin{itemize}
\item maximize $k_{e}$
\item choose $\delta\nu_{c}/2<300\mbox{ kHz}$ (see Sec.$\,$\ref{subsec:Cavity-Selection})
\item make $\tilde{y}_{5}$ symmetric about $\tilde{y}_{5}=0$
\end{itemize}
We can now write a more detailed expression for the laser frequency
detuning
\begin{equation}
y_{1}=\frac{\hat{\alpha}}{1+\hat{\alpha}}\left(\frac{n_{1}}{\hat{\alpha}}-\frac{n_{4}}{k_{e}\hat{C}}\right)\label{eq:locked-freq}
\end{equation}
where $\hat{\alpha}=\hat{\alpha}_{\mathrm{slow}}+\hat{\alpha}_{\mathrm{fast}}=\hat{H}_{\nu_{c}}\hat{K}_{\mathrm{slow}}\hat{G}_{\mathrm{slow}}\hat{T}+\hat{H}_{\nu_{c}}\hat{K}_{\mathrm{fast}}\hat{G}_{\mathrm{fast}}\hat{T}$
(see the loop in Fig.$\:$\ref{fig:setups}). As expected laser noise
$n_{1}$ is suppressed by the loop gain $1+\hat{\alpha}$ while noise
$n_{4}$ is only suppressed by $k_{e}\hat{C}$.

Increasing $P_{\mathrm{PD}}$ improves $k_{e}$ and SNR, but use caution
as it can have adverse consequences including heating of mirror coatings,
cavity drift \citep{MOGlabs19Doc}, error signal asymmetry \citep{Fox03NIST},
and cross-talk in a multi-color PDH setup. Finally, maximize $k_{e}$
by adjusting the phase-delay between the LO and mixer (Fig 4(a)).
We discuss choice of $\delta\nu_{c}$ in Sec.$\,$\ref{subsec:Cavity-Selection}.

Since $\hat{H}_{\nu_{c}}$ is linear only for $\tilde{y}_{1}\in(-\delta\nu_{c}/2,\delta\nu_{c}/2)$,
care must be exercised during operation to ensure this limit is not
violated. A proper error signal is symmetric about $\tilde{y}_{5}=0$.
The simplest asymmetry is an offset to the entire error signal. If
the offset is slowly varying, the result is a shift in the lock point
and reduced discriminator dynamic range. We note that the offset can
vary due to RAM \citep{Zhang14ol,Bi19ao}\footnote{RAM can be due to polarization drift upstream of the EOM and etalon
effects due to interference of stray reflections along the beam path
\citep{Zhang14ol,Fox03NIST}.}.

The transfer function $\hat{H}_{\nu_{c}}=\hat{D}\hat{P}k_{e}\hat{C}$
is straightforward to determine. $\hat{C}$ is determined by cavity
ring-down. The slope $k_{e}$ is best determined empirically. Increase
$|\hat{\alpha}|$ by raising loop filter gain $K_{P}$ so the error
signal $\tilde{y}_{5}$ oscillates well into saturation. Note the
peak to peak amplitude $\tilde{y}_{5}^{\mathrm{PP}}$ on an oscilloscope
and calculate $k_{e}=2\tilde{y}_{5}^{\mathrm{PP}}/\delta\nu_{c}$.
As a first-order approximation, $\hat{D}$ is in many situations dominated
by the low-pass filter and $\hat{P}=1$ in the limit of high PD bandwidth.
Precisely, $\hat{D}\hat{P}$ can be measured with a VNA as mentioned
in Appendix$\:$\ref{appendix:lock-in-detection}.

\subsection{Loop filter}

The loop filter provides frequency-dependent gain and phase advance
that can be tuned to optimize closed loop performance (see Sec.$\,$\ref{subsec:Loop-Components}).
We focus on analog PID loop filters which have several salient properties.
\begin{itemize}
\item adjustable fast-branch $\hat{K}_{\mathrm{fast}}$ parameters: $K_{P},f_{I},f_{D}$
as shown in Eq.$\:$\ref{eq:PID}
\item adjustable slow-branch $\hat{K}_{\mathrm{slow}}$ parameter: $f_{I}^{\mathrm{slow}}$,
$\hat{K}_{\mathrm{slow}}=-jf_{I}^{\mathrm{slow}}/f$
\item adjustable input offset $m_{6}^{\mathrm{DC}}$
\item propagation delay
\item low-frequency integral gain limit
\end{itemize}
As discussed in the introduction, the widely-used PID loop filter
provides 3 free parameters: $f_{I}$, $f_{D}$, and $K_{P}$ \citep{Bechhoefer05rmp}.\footnote{Note that some loop filters have adjustable parameters $K_{P},K_{I},K_{D}$
where the transfer function is $\hat{K}=K_{P}-jK_{I}/f+jK_{D}f$.
This is a poor choice since the pole is $f_{I}=K_{I}/K_{P}$ and the
zero is $f_{D}=K_{P}/K_{D}$ so the they can't be set independent
of $K_{P}$. Loop filters like this cost extra tuning work in optimizations.} The corresponding transfer function is Eq.$\:$\ref{eq:PID}. The
input offset $m_{6}^{\mathrm{DC}}$ can compensate for the error signal
offset. The propagation delay of a good analog PID can be as low as
$10\mbox{ ns}$ \citep{FALCpro,no-endorse}. This corresponds to less
than $-36^{\circ}$ at $10\mbox{ MHz}$.

Although $|\hat{\alpha}|\rightarrow\infty$ seems like it might be
optimal, it is common to limit the integral gain at low frequencies.
When initially engaging the lock, the error signal may be far from
zero due to laser noise and large I-gain near DC can drive the feedback
to saturation. This can be mitigated by first establishing an initial
lock using the fast branch and later slow branch. Note that even when
system performance is shot-noise limited, high-gain at low frequency
is not needed \citep{Mor97ijqe,wang2024pdh-lockable,Day92ijqe,Black00ajop}.

Digital loop filters based on field-programmable gate array (FPGA)
have been developed and successfully applied in AMO experiments including
laser locking \citep{Leibrandt15arX,Yu18rosi,Tourigny-Plante18rsi,Luda19rsi,Wiegand22rsi,Neuhaus24rosi}.
They can provide more diverse transfer functions with features like
providing more than $90^{\circ}$ lead or lag, multiple inputs and
outputs, auto-relock, integrator hold, and system discovery. However,
their delay is typically $>200\,\mathrm{ns}$ which ultimately limits
the phase lead and they are limited by high input noise $50-100\,\mathrm{nV/\sqrt{Hz}}$
at $1\,\mathrm{kHz}$ \citep{Yu18rosi,Leibrandt15arX}. Both are approximately
$20\text{-times}$ higher than those for a high-performance analog
loop filter \footnote{Commercial examples include the Toptica FALCpro and Vescent D2-125
\citep{no-endorse}.} In the case of laser linewidth reduction, maximizing $f_{\mathrm{UG}}$
is the highest priority and analog feedback is best for the fast branch.
For the slow-branch $\hat{K}_{\mathrm{slow}}$, delay is not a problem
and a digital loop filter may be a good choice \citep{Nagourney14OUP,Ryou17rosi}.
Additional resources for loop filter design include Ref.$\,$\citep{Hall99MP,Fox03NIST,Bechhoefer05rmp}.

\subsection{Cavity selection\protect\label{subsec:Cavity-Selection}}

\begin{figure}[ht]
\includegraphics[width=1\columnwidth]{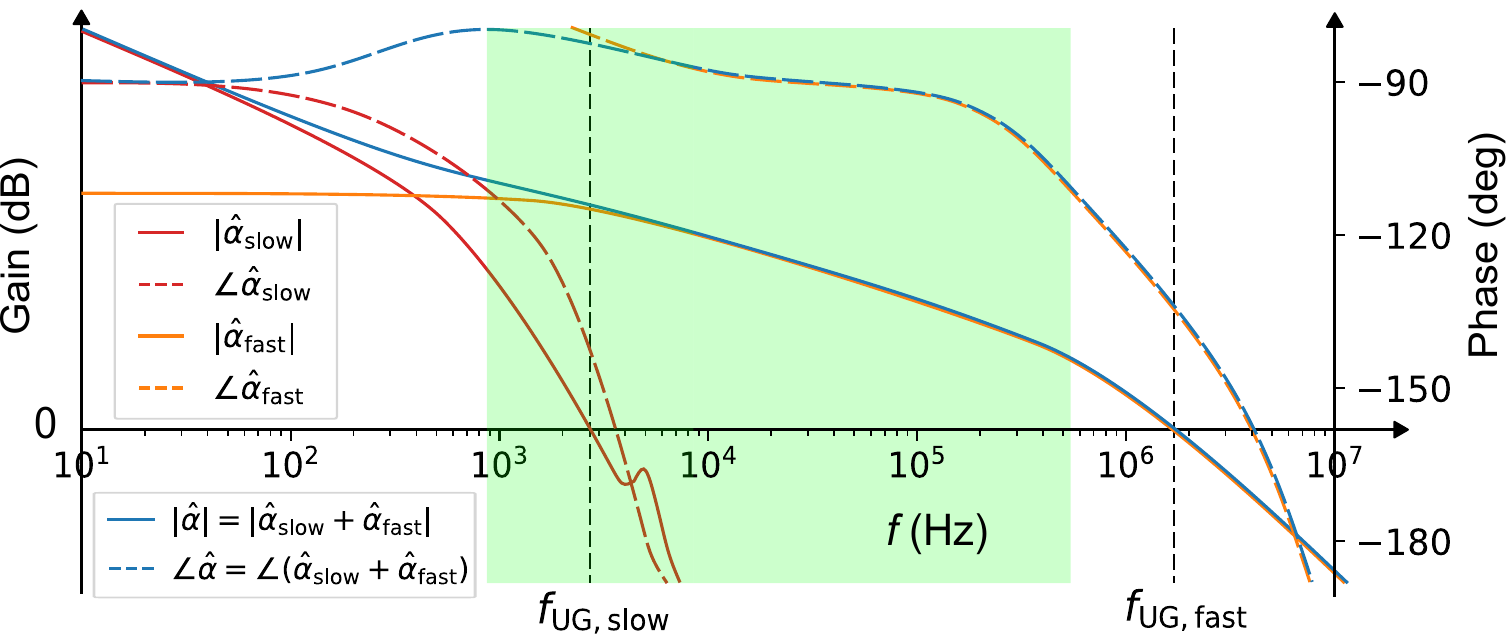} \caption{\protect\label{fig:cavity-selection}Sample Bode plots of the transfer
function of the fast ($\hat{\alpha}_{\mathrm{fast}}$) and slow branches
($\hat{\alpha}_{\mathrm{slow}}$). $f_{\mathrm{UG,slow}}$ and $f_{\mathrm{UG,fast}}$
are UG points for the slow- and fast-branch feedback. Our recommendation
is that the cavity pole $\delta\nu_{c}/2$ lies in green-shaded region.
}
\end{figure}

Finally we discuss is the linewidth $\delta\nu_{c}$ choice of the
reference cavity. Industry-standard reference cavities have an FSR
of $1.5\text{ GHz}$ leaving finesse (linewidth $\delta\nu_{c}$)
as the primary free parameter. Our recommendation balances the trade-offs
discussed below (Fig.$\:$\ref{fig:cavity-selection}).

A low-noise lock favors small $\delta\nu_{c}$. The PSD of Eq.$\:$\ref{eq:locked-freq}
is
\begin{equation}
S_{y1}=\frac{1}{\left|1+\hat{\alpha}\right|^{2}}S_{n1}+\frac{\left|\hat{\alpha}\right|^{2}}{\left|1+\hat{\alpha}\right|^{2}}\frac{1}{k_{e}^{2}|C|^{2}}S_{n4}\label{eq:Sy1_PSD}
\end{equation}
Since $1/(k_{e}^{2}|C|^{2})\propto\delta\nu_{c}^{2}+4f^{2}$, smaller
$\delta\nu_{c}$ gives less sensitivity to discriminator noise $n_{4}$.
Note that choosing $\delta\nu_{c}$ too small can make cavity alignment
hard because of low transmission and make locking hard since the laser
is commonly servo'd manually to $y_{1}\in(-\delta\nu_{c}/2,\delta\nu_{c}/2)$
before activating the lock.

The optimal choice of the cavity pole $\delta\nu_{c}/2$ to maximize
$f_{\mathrm{UG}}$ is $\delta\nu_{c}/2>\sqrt{10}f_{\mathrm{UG,fast}}$,
as indicated by the phase response of a low-pass filter (see Fig.$\:$\ref{fig:basics}(d)).
However, as shown in Eq.$\:$\ref{eq:Sy1_PSD}, a small $\delta\nu_{c}$
suppresses discriminator noise. Therefore, the sub-optimal but noise-favorable
choice is $\delta\nu_{c}/2<f_{\mathrm{UG,fast}}/\sqrt{10}$. Avoid
overlap of $\delta\nu_{c}/2$ with $f_{\mathrm{UG,fast}}$, else the
phase margin is more sensitive to laser power or RAM drift because
of steeper $d\angle\hat{\alpha}/df$ near $f_{\mathrm{UG,fast}}$.
If fast branch cannot fully suppress slow branch servo bump, less
$\angle\hat{C}$ when $\delta\nu_{c}/2>\sqrt{10}f_{\mathrm{UG,slow}}$
is helpful. For a typical ECDL we have $f_{\mathrm{UG,fast}}\sim1-2\text{ MHz}$
and $f_{\mathrm{UG,slow}}\sim1-10\text{ kHz}$, for which the recommended
range of $\delta\nu_{c}$ is the green-shaded range in Fig.$\:$\ref{fig:cavity-selection}.

State-of-the-art mirror coating can achieve a finesse of $10^{6}$
\citep{Jin22o,Guo22sa,Zhao10oc,Matei17prl}. Given a cavity length
of 10 to $50\text{ mm}$ \citep{Ludlow15rmp,Alnis08pra,Leibrandt11oe,Hafner15ol,Matei17prl},
a practical lower bound on linewidth is around $1\text{ kHz}$.

\section{Closed-loop optimization}

In an all-electronic system, a vector network analyzer (VNA) can be
used to probe each system component separately to determine its transfer
function and noise. The present context is a mixed optical-electronic
system so we in addition need an optical frequency discriminator (OFD).
Required properties of this OFD include sensitivity at the wavelength
of interest, a bandwidth well above the target $f_{\mathrm{UG}}$
and sensitivity to small frequency deviations (and therefore high
stability over time). We also want a frequency-agile optical source
such as a tunable laser or EOM. However, in many labs the only suitable
OFD available is the PDH discriminator $\hat{H}_{\nu_{c}}$ we want
to optimize! Further, measuring the PDH OFD transfer function by open-loop
techniques is impractical since the reference cavity linewidth is
typically small relative to the free-running laser linewidth. In this
chicken-and-egg situation the most expedient solution is to measure
what we can using open loop techniques, then rely on closed-loop measurements
to estimate the OFD and laser transfer functions. The closed loop
probe relies on a VNA to inject stimulus at summing points $m_{k}$
while measuring electrical loop signals $y_{k}$ which are related
as 
\begin{equation}
\begin{pmatrix}y_{5}\\
y_{6}\\
y_{8}
\end{pmatrix}=\frac{1}{1+\hat{\alpha}}\begin{pmatrix}\hat{H}_{\nu_{c}} & \hat{\alpha} & \frac{\hat{\alpha}_{\mathrm{fast}}}{\hat{K}_{\mathrm{fast}}}\\
-\hat{H}_{\nu_{c}} & 1 & -\frac{\hat{\alpha}_{\mathrm{fast}}}{\hat{K}_{\mathrm{fast}}}\\
-\frac{\hat{\alpha}_{\mathrm{fast}}}{\hat{G}_{\mathrm{fast}}} & \hat{K}_{\mathrm{fast}} & 1+\hat{\alpha}_{\mathrm{slow}}
\end{pmatrix}\begin{pmatrix}m_{2}\\
m_{6}\\
m_{8}
\end{pmatrix}\label{eq:closed-loop-relations}
\end{equation}
See signals $y_{k}$ and modulation $m_{k}$ in Fig.$\:$\ref{fig:setups}.
Then we are in a position to optimize the overall loop.

\subsection{System optimization workflow}

\label{sec:system-optimization-routine}

\begin{figure}
\centering\includegraphics[width=1\columnwidth]{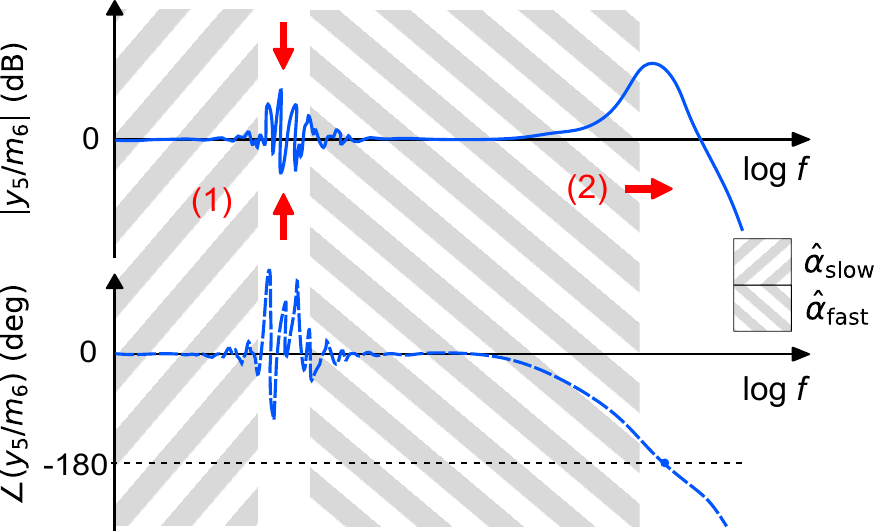}

\caption{\protect\label{fig:optimization}A sample closed-loop transfer function
$y_{5}/m_{6}$ for an incorrectly tuned system. We want to adjust
the loop parameters to (1) obtain $y_{5}/m_{6}\approx1$ for $f<f_{\mathrm{180}}^{\mathrm{CL}}/10$.
(2) maximize $f_{\mathrm{180}}^{\mathrm{CL}}$. The gray-white striped
regions show the ranges in which the slow and fast feedback attenuate
noise. We aspire to eliminate the white gap between them.}
\end{figure}

We use the following workflow to tune loop filter parameters and check
for several misconfiguration errors. It consists of four steps: prepare,
measure, optimize and check.
\begin{description}
\item [{prepare}]~
\end{description}
\begin{enumerate}
\item Measure the transfer function of each loop component.
\item Sweep the laser frequency across the cavity resonance and observe
the error signal $\tilde{y}_{5}$ line shape on an oscilloscope (Fig.$\:$\ref{fig:error-drift}).
Make the error signal symmetric about $\tilde{y}_{5}=0$ by adjusting
the phase $\phi_{o}$ of the RF source (Fig.$\:$\ref{fig:setups}(a)).
\item Sweep across the cavity resonance and observe the fast loop filter
output $\tilde{y}_{8}$ on an oscilloscope. Adjust the input offset
$m_{6}^{\text{DC}}$ to make $\tilde{y}_{8}$ symmetric about zero.
\item To null the slow loop offset do as follows. Far detune the laser from
resonance. Turn on the slow loop filter and observe the rate at which
output $\tilde{y}_{7}$ rails. Adjust the slow input offset to minimize
the rate. This offset is not shown in Fig.$\:$\ref{fig:setups}.
\item To initialize the loop filter: minimize $f_{I}$ and disable $f_{D}$.
\item Close the loop and obtain a weak lock starting with the fast branch
of the loop filter. Increase $K_{P}$ until the loop oscillates then
reduce $K_{P}$ by 50\%. Next, add $f_{I}$ until the loop oscillates
then reduce $f_{I}$ by 50\%. Finally add $K_{P}$ back a little.
Signals $\tilde{y}_{5}$ and $\tilde{y}_{8}$ show periodic waveform
on the oscilloscope when the loop oscillates.
\item Next enable the slow integrator $\hat{K}_{\mathrm{slow}}$. Increase
$f_{I}^{\text{slow}}$ until the loop oscillates. Then reduce it until
oscillation disappears.
\end{enumerate}
\begin{description}
\item [{measure}]~
\end{description}
\begin{enumerate}
\item Measure $y_{5}/m_{6}$ by connecting a VNA as shown in Fig.$\:$\ref{fig:setups}(a).
Program the VNA to apply stimulus at $m_{6}$ and record the response
at $y_{5}$ for frequencies up to $10\text{ MHz}$. The injected amplitude
should be low enough to keep $\tilde{y}_{5}$ in the linear region;
vary the amplitude of $m_{6}$ with frequency as needed. Make a Bode
plot of $y_{5}/m_{6}$.
\item Take note of the frequency $f_{\mathrm{180}}^{\mathrm{CL}}$ which
is where $\angle(y_{5}/m_{6})=-180^{\circ}$ \footnote{Observe that $f_{\mathrm{180}}^{\mathrm{CL}}=f_{\mathrm{180}}$. This
is because $\hat{\alpha}(f_{\mathrm{180}})$ is real and $-1<\hat{\alpha}(f_{\mathrm{180}})<0$
when the loop is stable, so too $\hat{\alpha}/(1+\hat{\alpha})$ is
real and negative. Thus $f_{\mathrm{180}}^{\mathrm{CL}}=f_{\mathrm{180}}$.
Checking $f_{\mathrm{180}}^{\mathrm{CL}}$ is quicker. Improving $f_{\mathrm{180}}$
is equivalent to improve $f_{\mathrm{UG}}$.}.
\item Measure $\tilde{y}_{5}$ on an oscilloscope and note its RMS value
$\tilde{y}_{5}^{\mathrm{RMS}}$ which is proportional to laser frequency
fluctuation $\tilde{y}_{1}^{\mathrm{RMS}}$ \footnote{Note that $\tilde{y}_{5}^{\mathrm{RMS}}=\sqrt{\int_{0}^{\infty}S_{y5}df}=\sqrt{\int_{0}^{\infty}\left|\hat{H}_{\nu_{c}}\right|^{2}S_{y1}df}=\left|\hat{H}_{\nu_{c}}(\xi)\right|^{2}\sqrt{\int_{0}^{\infty}S_{y1}df}=\left|\hat{H}_{\nu_{c}}(\xi)\right|^{2}\tilde{y}_{1}^{\mathrm{RMS}}$,
where $\xi$ is a finite frequency.}. Equivalently, you can avoid $1/f$ noise in $\tilde{y}_{5}$ by
using a spectrum analyzer to measure $S_{y4}$.
\end{enumerate}
\begin{description}
\item [{optimize}]~
\end{description}
First, check if there is excess low-frequency noise; we want $|y_{5}/m_{6}|\approx1$
and $\angle y_{5}/m_{6}\approx0^{\circ}$ for $f<f_{\mathrm{180}}^{\mathrm{CL}}/10$
(see Fig.$\:$\ref{fig:optimization}). If not, reduce fast-branch
high-pass $f_{\mathrm{HP,fast}}$. If the problem persists, reduce
slow-branch $f_{I}^{\text{slow}}$.

Next, try to increase $f_{\mathrm{180}}^{\mathrm{CL}}$.
\begin{enumerate}
\item Enable and reduce $f_{D}$ until oscillation, then raise it by 50\%.
Increase $K_{P}$ until oscillation, then reduce it by 50\%. Increase
$f_{I}$ until oscillation, then reduce it by 50\%. Then, reduce $f_{D}$
until oscillation, then raise it by 50\%. Loop over adjustments to
$K_{P}$, $f_{I}$, and $f_{D}$ repeatedly until no further improvements
can be made. Finally, slightly raise $K_{P}$ again.
\item While observing $\tilde{y}_{5}^{\mathrm{RMS}}$, minimize the observable
while fine tuning $K_{P}$. Alternately, you may minimize $\sqrt{\int_{a_{-}}^{a_{+}}S_{y4}df}$,
where $a_{\pm}=\Omega/2\pi\pm5\text{ MHz}$\footnote{Integrate over a fixed range larger than larger $f_{\mathrm{180}}^{\mathrm{CL}}$,
here we use $5\text{ MHz}$.} 
\end{enumerate}
\begin{description}
\item [{check}]~
\end{description}
Start by calculating the open-loop transfer function $\hat{\alpha}$
from the VNA measurement. 
\begin{equation}
\frac{y_{5}}{m_{6}}=\frac{\hat{\alpha}}{1+\hat{\alpha}}\implies\hat{\alpha}=\frac{y_{5}/m_{6}}{1-y_{5}/m_{6}}\label{eq:closed-to-open}
\end{equation}
Generate a Bode plot of $\hat{\alpha}$ and locate $f_{\mathrm{UG}}$
and $\phi_{m}$. Then check the following.
\begin{enumerate}
\item If $f_{\mathrm{UG}}$ is not as high as you hoped, revisit Sec.$\:$\ref{sec:PDH-Lock-Component-Choices}
and use components with less phase lag. Adjusting the loop filter
won't help.
\item If $\tilde{y_{5}}^{\mathrm{RMS}}$ (or $S_{y4}$) is satisfactory
but $\phi_{m}<30^{\circ}$, you can increase $\phi_{m}$ at the expense
of degraded $\tilde{y}_{5}$ RMS by reducing $K_{P}$. Increased $\phi_{m}$
improves loop stability.
\item We expect that $\angle\hat{\alpha}$ is the sum of the phase shift
from the individual components: $\angle\hat{K}_{\mathrm{fast}}+\angle\hat{G}_{\mathrm{fast}}+\angle\hat{C}+\angle\hat{D}+\angle\hat{P}+\angle\hat{T}$.
If excess phase lag is observed, dig to find the source.
\end{enumerate}

\subsection{Example optimization}

\label{sec:example-optimization}

\begin{figure*}
\centering \includegraphics[width=0.7\paperwidth]{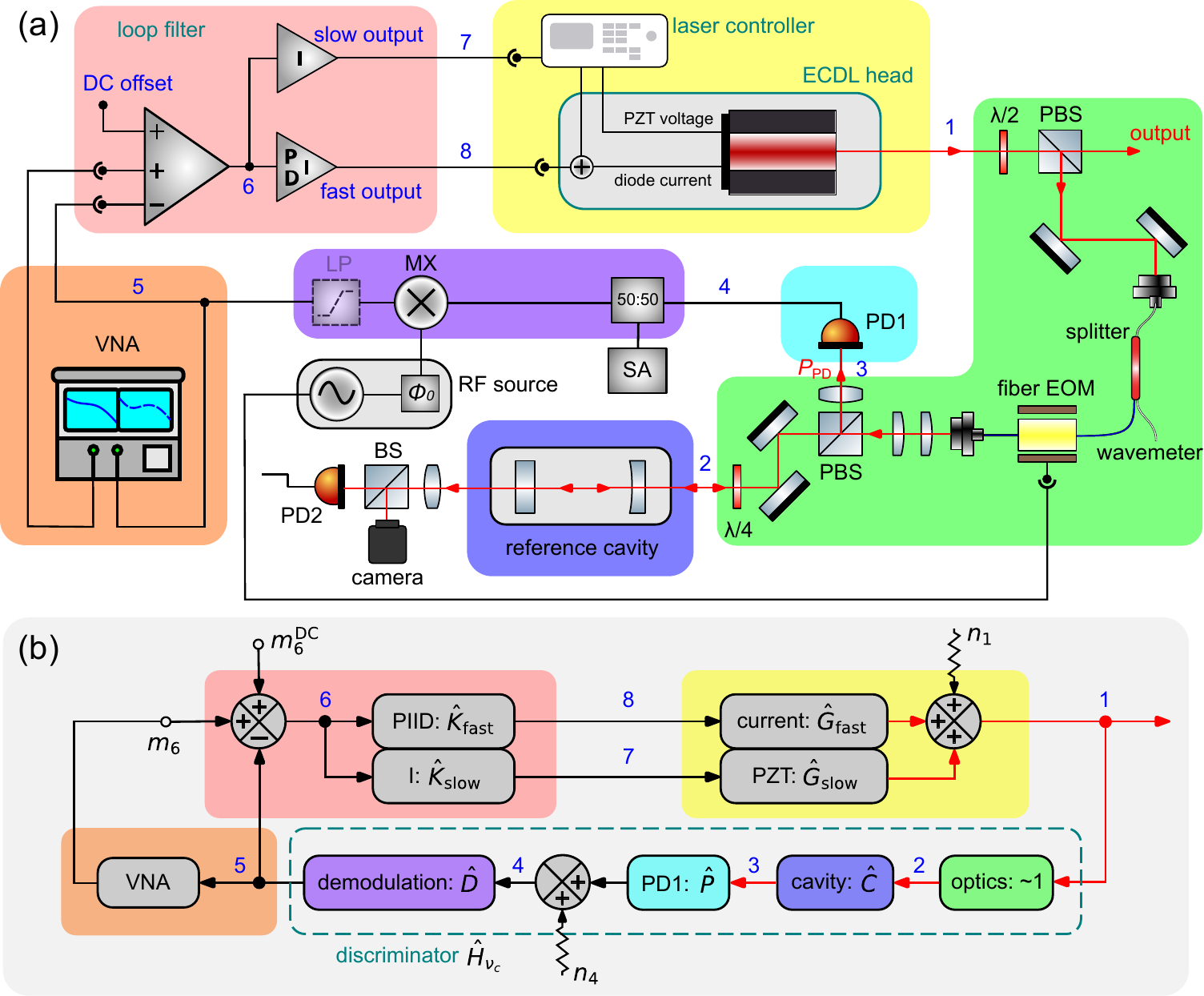}\caption{\protect\label{fig:setups}(a) A schematic for the components in a
PDH laser frequency stabilization setup. Positions within the loop
are labeled with number $k$ in blue. Loop signals $y_{k}$ is defined
to include modulation $m_{k}$ or noise $n_{k}$ at summing points
(not drawn for $m_{2}$ and $m_{8}$). Black (red) labels denote electrical
(optical) signals. Light from an ECDL acquires sidebands at $\pm\Omega$
from a fiber EOM, is reflected off a reference cavity and incident
on a photodetector (PD1). The PD1 signal is demodulated to generate
error signal $y_{5}$. In closed-loop analysis, a VNA injects stimulus
at position 6. Outputs of the loop filter are sent to the laser actuators,
completing the loop. A camera and PD2 are used to check cavity transmission
and alignment. A spectrum analyzer (SA) is at position $4$. (b) A
block diagram of the feedback loop in (a). Transfer functions for
each component are denoted by capital letters (eg $\hat{C}$). A dashed
line encloses the optical frequency discriminator $\hat{H}_{\nu_{c}}=\hat{D}\hat{P}k_{e}\hat{C}$.
Free-running laser noise $n_{1}$ and discriminator noise $n_{4}$
(including the noises in PD, RAM noise, cavity drifts, etc.) are summed
into the loop as indicated.}
\end{figure*}

In this section we illustrate application of the system optimization
workflow (Sec.$\,$\ref{sec:system-optimization-routine}) to improve
$\hat{\alpha}$ for a representative laser system. There are three
system configurations (Table \ref{tab:freq_points}) where the loop
is diagnosed and incrementally improved. The optical and electronic
elements of the PDH setup are shown in Fig.$\:$\ref{fig:setups}.

\subsubsection{configuration 1}

The initial, unoptimized configuration is as follows. The laser is
a 1650 nm Littrow ECDL with a $70\,\mathrm{dB}$ optical isolator.
We observe a free-running frequency fluctuation of $\sim200\text{ kHz}$
over a $10\text{ ms}$ integration window by monitoring side-of-fringe
transmission from a test cavity with a full-width at half-max (FWHM)
linewidth $4.2\,\text{MHz}$ (Thorlabs SA200-12B \citep{no-endorse}).
The laser controller has external inputs to modulate the PZT voltage
(slow branch) and the laser diode current (fast branch). The PZT actuator
has a lowest mechanical resonance at $6\,\mathrm{kHz}$. Using the
technique in Appendix \ref{appendix:laser-frequency-modulation} we
observe that current modulation exhibits a $-52^{\circ}$ phase shift
at $1\,\mathrm{MHz}$.

Sidebands are applied to the laser light using a fiber EOM (iXblue
MPZ-LN-10 \citep{no-endorse}) driven at $\Omega/2\pi=20\text{ MHz}$
with voltage $\beta/\pi\cdot V_{\text{\ensuremath{\pi}}}\approx1.5\text{ V}$
for $\beta=1.082$. The thermally and acoustically isolated reference
cavity (Stable Laser Systems \citep{SLS-cavity,no-endorse}) is made
of ultra-low expansion glass (ULE). The cavity has planar and concave
($R=-50\,\mathrm{cm}$) mirrors separated by $10\text{ cm}$. It is
temperature-stabilized at $24.7\mathrm{\,^{\circ}C}$ where the linear
thermal expansion coefficient is zero. The cavity FSR is $1.5\text{ GHz}$,
its FWHM as measured by ring down is $\delta\nu_{c}=45.7\,\mathrm{kHz}$
(see Appendix$\:$\ref{appendix:ringdown}) and its finesse at $1650\text{ nm}$
is $\mathcal{F}=\nu_{\mathrm{FSR}}/\delta\nu_{c}\approx33,000$. The
laser was consistently locked to the same cavity transmission feature
at $\nu_{c}=181.65917\mbox{ THz}$ using a Michelson interferometer-based
wave meter (Bristol 228A \citep{no-endorse}) to disambiguate cavity
transmission features.

The PD for cavity-reflected light (PD1 in Fig.$\:$\ref{fig:setups}(a))
is a reverse-biased InGaAs diode (Thorlabs PDA10D2 \citep{no-endorse})
with an $\text{NEP}=39\text{ pW/}\sqrt{\text{Hz}}$, a bandwidth $f_{\mathrm{PD}}=20\text{ MHz}$,
and $\angle\hat{P}=-6.2^{\circ}$ at $1\,\mathrm{MHz}$. An optical
power of $850\,\mathrm{\mu W}$ incident on the cavity results in
$650\,\mathrm{\mu W}$ at PD1. The SNR for PD1 is greater than $2700$.
We use a Level 7 mixer (MiniCircuits ZAD-1-1+ \citep{no-endorse}).
Using the technique in Appendix \ref{appendix:lockin-transfer-function}
we determine that the mixer adds a phase lag of $-1^{\circ}$ at $1\,\mathrm{MHz}$.
A dual-channel RF signal generator (Rigol DG822 \citep{no-endorse})
generates $\Omega$ with a phase offset for the tone sent to the mixer
LO port. The low-pass (LP) (Minicircuits BLP-10.7+ \citep{no-endorse})
is an 8th order modified Butterworth-type filter with a corner frequency
of $14\text{ MHz}$ and a $-22^{\circ}$ phase lag at $1\,\mathrm{MHz}$;
the LP provides $-34\text{ dB}$ suppression at $\Omega$. A high-speed
analog loop filter was used with $<45^{\circ}$ electronic phase delay
at $10\text{ MHz}$ (Toptica FALC 110 \citep{FALC110,no-endorse}).

We used a vector network analyzer (VNA, OMICRON Lab Bode 100 \citep{no-endorse})
for transfer function measurements and a spectrum analyzer (Rohde
\& Schwarz FSV40-N \citep{no-endorse}) to measure the noise spectrum.

\begin{figure*}
\centering \includegraphics[width=0.8\paperwidth]{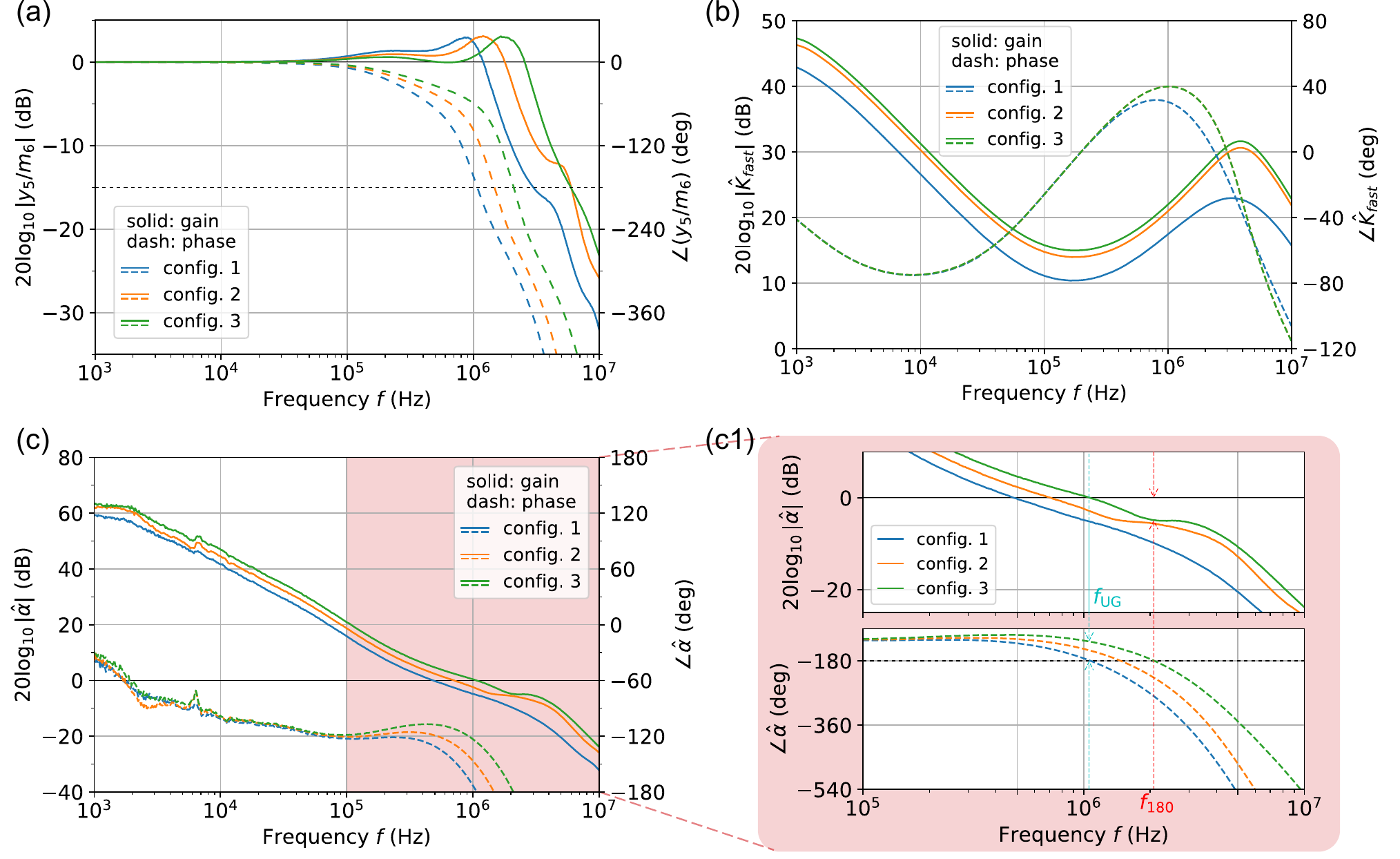}\caption{\protect\label{fig:config-measurements}Here we show performance in
configuration 1, 2, and 3. (a) Bode plots of closed-loop transfer
functions $y_{5}/m_{6}$ for the three configurations. The VNA resolution
bandwidth is $10\,\mathrm{Hz}$. (b) Bode plots of the fast-branch
loop filter $\hat{K}_{\mathrm{fast}}$. (c) Bode plots of open-loop
transfer function $\hat{\alpha}$ calculated from $\frac{y_{5}/m_{6}}{1-y_{5}/m_{6}}$.
As expected the slope of $|\hat{\alpha}|$ is approximately $-20\text{ dB}/\text{decade}$.
(c1) Zoom-in of the red-shaded area of plot (c). For configuration
3, we call out the UG point $f_{\text{UG}}$ and phase crossover point
$f_{180}$.}
\end{figure*}

\subsubsection{configuration 2}

We were not satisfied with the low $f_{\mathrm{UG}}\approx0.5\text{ MHz}$
for configuration 1. We found several components could be optimized
to reduce their phase lags. The following changes were made for configuration
2. 

The PD1 was changed to PDA05CF2 \citep{no-endorse} $(f_{\mathrm{PD}}=150\mbox{ MHz})$
which has lower noise $\mathrm{NEP}=6.3\,\mathrm{pW/\sqrt{Hz}}$ (see
Appendix \ref{appendix:lockin-transfer-function}) at $1650\,\mathrm{nm}$
and smaller phase lag $\angle\hat{P}=-1.3^{\circ}$ at $1\,\mathrm{MHz}$.
The optical power $P_{\mathrm{PD}}$ was reduced from $650\,\mathrm{\mu W}$
to $430\,\mathrm{\mu W}$ to accommodate increased PD efficiency,
and provides a better SNR$\approx6800$. The loop length was reduced
by $6.8\,\mathrm{m}$. More phase lead was added by increasing the
D term in the loop filter. In total, we get $+26.6^{\circ}$ phase
lead at 1 MHz relative to configuration 1.

\subsubsection{configuration 3}

We found the discrete low-pass in the demodulation circuit significantly
contributed to phase lag. Removing it results in reduced SNR, but
this is acceptable for us. Instead, we relied on the internal low-pass
behavior of the loop filter. We get $+23.5^{\circ}$ phase lead at
1 MHz relative to configuration 2. The bulk of the remaining phase
lag at 1 MHz is due to laser diode fast current modulation path. Now
we found no optimizations for our setup using the workflow.

\subsubsection*{optimization results}

Table$\:$\ref{tab:freq_points} summarizes the performance for each
configuration. We see improvements in $f_{\mathrm{UG}}$ while $\phi_{m}$
remains in the recommended range. In Table$\:$\ref{tab:phase} we
show that from measurement of the individual transfer functions, we
can account for all phase delays in the loop. This gives us confidence
that we understand the origin of phase lags in our setup and have
addressed them during optimization.

\begin{figure*}
\includegraphics[width=0.7\paperwidth]{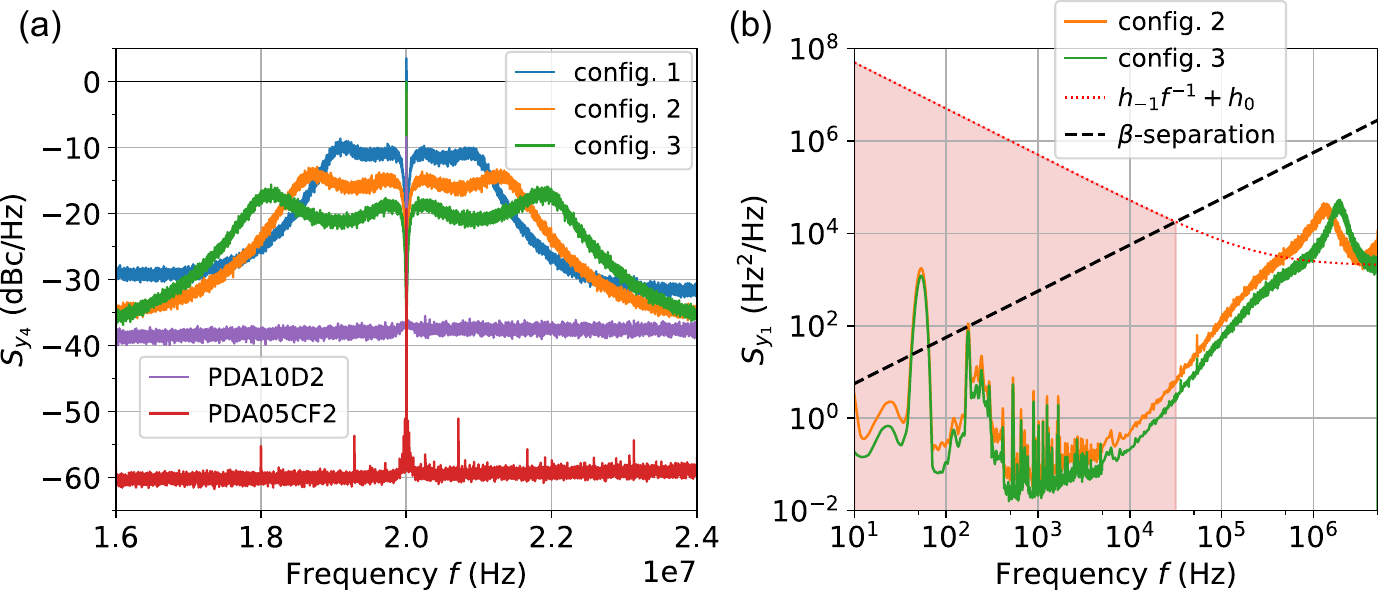} \caption{\protect\label{fig:spectra}(a) \lyxadded{JWB}{Sat Dec 21 00:16:16 2024}{}The
power spectral density $S_{y4}$ of the three configurations and the
two PDs. The PSD are normalized to configuration 3 and have a $1$
kHz resolution bandwidth. Photodetector baselines are measured with
the laser blocked; the $20\,\mathrm{MHz}$ carrier leakage is due
to imperfect mixer LO-RF isolation. (b) Here we show evidence of linewidth
narrowing. We plot the PSD $S_{y1}$ of the configuration 2 and 3
(solid lines) while locked and a model for the free-running laser
(dotted red line). We also plot the $\beta$-separation line (dashed
black). The PDH-locked $S_{y1}$ are calculated from measured $S_{y4}$
with a $10\text{ Hz}$ resolution bandwidth and the PD noise baseline
is subtracted.}
\end{figure*}

We also measured the noise spectra $S_{y4}(f)$ of cavity-reflected
signal $y_{4}$ by a spectrum analyzer (Fig.$\:$\ref{fig:spectra}(a)).
Servo bumps $f_{\mathrm{bump}}$ are summarized in Table$\,$\ref{tab:freq_points}.
Improving $f_{\mathrm{UG}}$ pushes servo bumps to higher frequencies
and enhances the noise suppression for all $f<f_{\mathrm{bump}}$.

\begin{table}[b]
\begin{tabular}{|c|c|c|c|}
\hline 
Configuration & UG point & phase margin & servo bump\tabularnewline
\hline 
 & $f_{\mathrm{UG}}$ (MHz) & $\phi_{m}$ (deg) & $f_{\mathrm{bump}}$ (MHz)\tabularnewline
\hline 
\hline 
1 & 0.49 & 50 & 0.91\tabularnewline
\hline 
2 & 0.71 & 51 & 1.32\tabularnewline
\hline 
3 & 1.06 & 54 & 1.94\tabularnewline
\hline 
\end{tabular}

\caption{\protect\label{tab:freq_points}A summary of key metrics for each
configuration.}
\end{table}

\begin{table}[b]
\begin{tabular}{|c|c|c|}
\hline 
Component & \begin{cellvarwidth}[t]
\centering
Phase shifts 

(deg) at 

$f_{\mathrm{UG}}=1.06\,\mathrm{MHz}$
\end{cellvarwidth} & Source\tabularnewline
\hline 
\hline 
$\angle\hat{K}_{\mathrm{fast}}$ & $+40$ & from Fig.$\,$\ref{fig:config-measurements}(b)\tabularnewline
\hline 
$\angle\hat{G}_{\mathrm{fast}}$ & $-52$ & see Appendix.$\,$\ref{appendix:laser-frequency-modulation}\tabularnewline
\hline 
$\angle\hat{C}$ & $-89$ & calculated from Eq.$\,$\ref{eq:cavity-transfer-function}$^{\text{[a]}}$\tabularnewline
\hline 
$\angle(\hat{D}\hat{P})$ & $-9$ & see Appendix$\,$\ref{appendix:lockin-transfer-function}\tabularnewline
\hline 
$\angle\hat{T}$ & $-15$ & $-360^{\circ}f_{\mathrm{UG}}\tau_{l}$$^{\text{[b]}}$\tabularnewline
\hline 
 &  & \tabularnewline
\hline 
$\angle\hat{\alpha}$ & $-126$ & from Fig.$\,$\ref{fig:config-measurements}(c1)\tabularnewline
\hline 
\end{tabular}

\caption{{\footnotesize\protect\label{tab:phase}}Tabulation of the configuration
3 phase shifts at $f_{\mathrm{UG}}=1.06\text{ MHz}$. The sum of the
phase shifts from individual components differs from the measured
$\angle\hat{\alpha}$ by less than $5^{\circ}$. Since $f_{\mathrm{UG}}=1.06\text{ MHz}$,
$\hat{K}_{\mathrm{slow}}(f_{\mathrm{UG}})\approx0$ and $\hat{G}_{\mathrm{slow}}(f_{\mathrm{UG}})\approx0$,
we ignore $\hat{\alpha}_{\mathrm{slow}}(f_{\mathrm{UG}})$. {\footnotesize Remarks:
{[}a{]} $\delta\nu_{c}=45.7\,\mathrm{kHz}$, see the ring-down measurement
in Appendix$\:$\ref{appendix:ringdown}. {[}b{]} Here $\tau_{l}$
is the delay due to signal propagation thru $2.1\,\mathrm{m}$ free
space, $4.9\,\mathrm{m}$ fiber, and $1.7\,\mathrm{m}$ coaxial cable.
Dispersion in the fiber is negligible for our monochromatic continuous
light.}}
\end{table}

The single-sided laser frequency PSD $S_{y1}$ is plotted in Fig.$\:$\ref{fig:spectra}(b).
This quantity is calculated by shifting $S_{y4}$ by $-\Omega/2\pi$
and dividing by $|\hat{H}_{\nu_{c}}/\hat{D}|^{2}$. The error signal
slope was measured to be $k_{e}=2.69\text{ mV/kHz}$. We use the $\hat{D}$
and $\hat{P}$ from Appendix$\:$\ref{appendix:lockin-transfer-function}.

We would like to compare $S_{y1}$ for the free-running and for the
locked laser, however we can't measure the free-running PSD due to
the narrow linewidth of our reference cavity. Instead we model the
free-running behavior by Eq.$\:$\ref{eq:laser-power-law}. Since
above $f_{\text{UG}}$ there is no noise suppression even when feedback
is active, the value of $h_{0}$ can be determined from the closed-loop
response: $h_{0}=S_{y_{1}}(5\text{ MHz})=2\times10^{3}\text{ Hz}^{2}/\text{Hz}$.
We set $h_{-1}=5\times10^{8}\text{ Hz}^{2}$ which is close to the
values from reported frequency noise spectrum of Toptica ECDLs \citep{DLpro,Schmidt-Eberle23Doc,Kolodzie24oe,Kervazo24arX}.
The shaded red area contributes to a linewidth of $\sim150\text{ kHz}$
according to the $\beta$-separation line theory \citep{Domenico10ao}.
In comparison with the free-running laser performance it's clear configuration
2 and 3 have substantially reduced laser linewidth.

We are satisfied with our laser performance after two iterations of
optimization where we pushed $f_{\mathrm{UG}}$ from $0.49$ to $1.06\text{ MHz}$
with the recommended phase margin and showed linewidth narrowing based
on an analysis of the PSD of $S_{y4}$. We also systematically checked
that the sum of phase shifts from all individual components, measured
in open loop, was in a good agreement with $\angle\hat{\alpha}$,
measured in closed loop. So we are confident in our understanding
of the sources of phase lag in our system. More detailed characterization
of the stability of our locked laser could be obtained using the delayed
self-heterodyne technique or beating with self-referenced optical
frequency comb.

\section{Summary}

We hope this tutorial proves useful to practitioners learning to implement
laser linewidth narrowing using the PDH technique. Relative to what's
been published before our treatment pulls together ideas on component
optimization and VNA informed loop filter optimization.

\paragraph{Acknowledgments}

We thank Dr. Alessandro Restelli at the University of Maryland, Dr.
David Nadlinger at the University of Oxford, and Dr. Chin-wen Chou
at the National Institute of Standards and Technology for their helpful
discussion. This work was funded by Army Research Laboratory (Sponsor
Award Number: W911NF2420107).

\paragraph{Author contributions}

W.W. developed the conceptual framework and methodology for the control
theory tutorial, component selection, and closed-loop optimization
sections. W.W. also conducted the experimental investigation and validation
presented in the manuscript, including data acquisition, processing,
and visualization. S.S. developed the conceptualization and methodology
for the closed-loop optimization section and shared the early-stage
investigation. J.B. contributed to the conceptualization of the control
theory tutorial and component selection sections, provided supervision,
managed the project administration, and secured resources and funding.
W.W. and J.B. wrote the primary manuscript, with contributions from
S.S. All authors reviewed the final manuscript.

\section{Appendix}

\subsection{Phase shift in lock-in detection}

\label{appendix:lock-in-detection} We derive the phase shift in demodulated
error signal $\tilde{y}_{5}$ due to the PD between the modulation
and demodulation. The light reflected from cavity generates a signal
$\tilde{y}_{4}$ on the PD. We check $\tilde{y}_{4}$ at Fourier frequency
$f$ ($2\Omega$ terms are removed by the low-pass filter), 
\[
\tilde{y}_{4}(t)=V_{-}\cos\left[(\Omega-2\pi f)t+\phi_{-}\right]+V_{+}\cos\left[(\Omega+2\pi f)t+\phi_{+}\right]
\]
$V_{\pm}$ and $\phi_{\pm}$ represent different gains and phase shifts
at $\Omega/2\pi\pm f$ added by the PD ($\phi=\angle\hat{L}_{\mathrm{PD}}$,
$\hat{L}_{\mathrm{PD}}$ is the PD transfer function without lock-in).
$\tilde{y}_{4}$ is then mixed with a LO signal with a tunable demodulation
phase $\phi_{0}$ to generate the error signal $\tilde{y}_{5}$, 
\begin{align*}
\tilde{y}_{5} & \propto\tilde{y}_{4}(t)V_{\mathrm{LO}}\cos\left(\Omega t+\phi_{0}\right)\\
 & \propto\frac{1}{2}V_{\mathrm{LO}}\sqrt{A^{2}+B^{2}}\cos\left(2\pi ft+\varphi\right)
\end{align*}
\begin{align*}
A & :=V_{-}\cos\left(\phi_{0}-\phi_{-}\right)+V_{+}\cos\left(\phi_{+}-\phi_{0}\right)\\
B & :=V_{-}\sin\left(\phi_{0}-\phi_{-}\right)+V_{+}\sin\left(\phi_{+}-\phi_{0}\right)
\end{align*}
\begin{align*}
\tan\varphi & =\frac{B}{A}=\frac{V_{-}\sin\left(\phi_{0}-\phi_{-}\right)+V_{+}\sin\left(\phi_{+}-\phi_{0}\right)}{V_{-}\cos\left(\phi_{0}-\phi_{-}\right)+V_{+}\cos\left(\phi_{+}-\phi_{0}\right)}
\end{align*}
A single $\phi_{0}$ cannot compensate both $\phi_{\pm}(\Omega/2\pi\pm f)$.
 Because the PD bandwidth  is usually much higher than $\Omega/2\pi$,
phase $\phi$ is approximately linear, so choose $\phi_{0}=\phi(\Omega/2\pi)\approx\frac{1}{2}(\phi_{-}+\phi_{+})$,
we get 
\begin{align}
\varphi(f) & =\left[\phi(\Omega/2\pi+f)-\phi(\Omega/2\pi-f)\right]/2
\end{align}
The gain of PD has been absorbed in the definition of slope $k_{e}$,
so we define PD transfer function in lock-in detection as $\hat{P}(f)=e^{i\varphi(f)}$.

\subsection{Cavity ring-down measurement}

\label{appendix:ringdown}
\begin{figure}
\includegraphics[width=0.9\columnwidth]{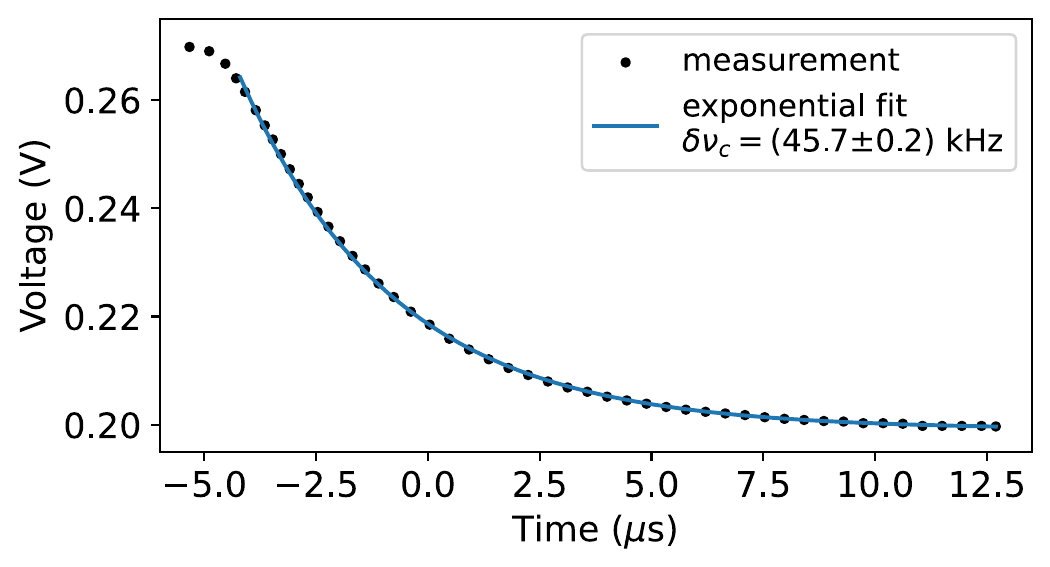}\caption{\protect\label{fig:ringdown}The reference cavity linewidth $\delta\nu_{c}$
is determined by a ring-down measurement as described in the text.
Initial transient behavior is excluded from the fit.}
\end{figure}
The reference cavity linewidth $\delta\nu_{c}$ was determined from
a ring-down experiment. The PD2 in Fig.$\:$\ref{fig:setups} measured
the power transmitted by the cavity. A wideband EOM amplitude modulator
(AM-EOM, iXblue MXER-LN-10) was inserted after the optical splitter.
The AM-EOM extinction was small but sufficient for this measurement.
A ring-down measurement cycle consisted of the steps: lock the laser
to the cavity, attenuate the laser light with the AM-EOM and record
the decaying PD2 signal. The ring-down trace in Fig.$\:$\ref{fig:ringdown}
is an average of 256 ring-down repetitions. An exponential fit to
this average yields $\delta\nu_{c}=(45.7\pm0.2)\:\text{kHz}$.

\subsection{Frequency modulation via diode current}

\label{appendix:laser-frequency-modulation}

\begin{figure}
\includegraphics[width=1\columnwidth]{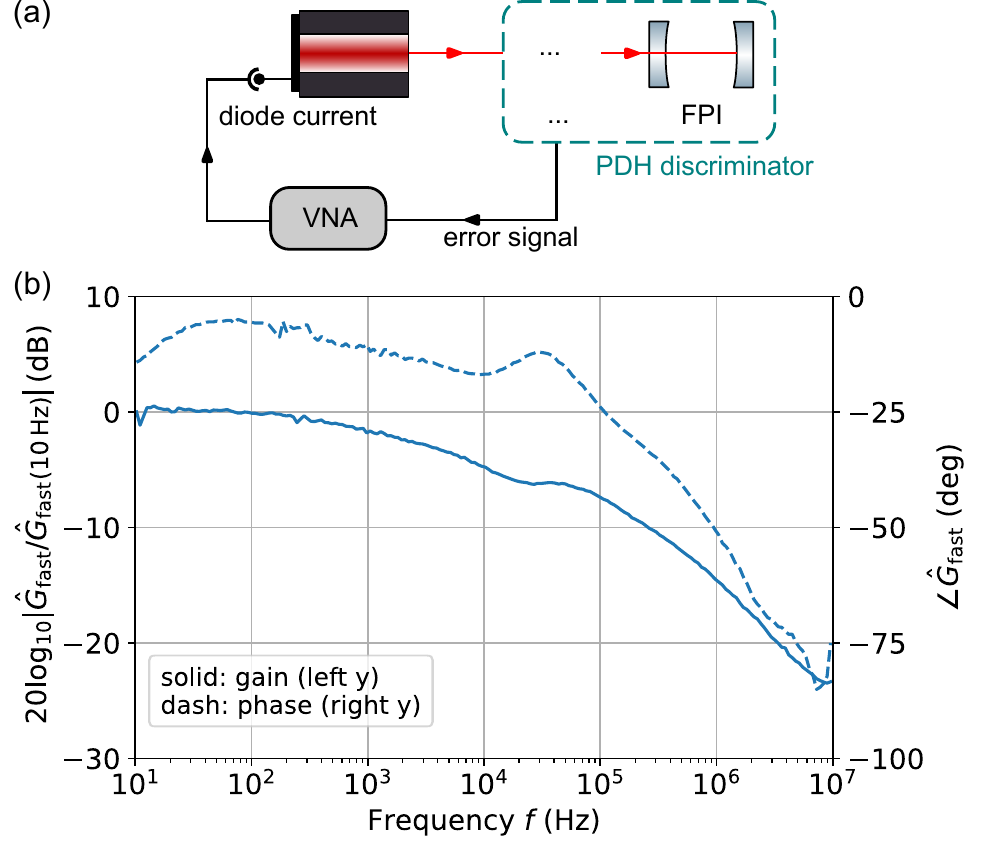}

\caption{\protect\label{fig:diode-mod_Bode}Setup (a) was used to measure the
transfer function $\hat{G}_{\mathrm{fast}}$ of the laser fast branch
actuator. (b) A Bode plot of the relative gain and phase for the fast
branch frequency modulation via diode current.}
\end{figure}

We measured the transfer function of the ECDL fast branch ($\hat{G}_{\mathrm{fast}}$)
using a PDH discriminator (Fig.$\:$\ref{fig:diode-mod_Bode}(a)).
The discriminator configuration is similar to Fig.$\:$\ref{fig:setups}(a),
except that the SLS optical cavity was replaced by a Fabry-Perot interferometer
(FPI, Thorlabs SA200-12B \citep{no-endorse}) with linewidth $\delta\nu_{\mathrm{FPI}}=5.4\text{ MHz}$\footnote{from an independent linewidth measurement}.
The FPI linewidth is broad relative to the free-running laser frequency
instability over the duration of our measurement. The transfer function
was measured by tuning the laser on resonance and monitoring the PDH
error signal while modulating the diode current using a VNA. The laser
transfer function $\hat{G}_{\mathrm{fast}}$ was determined after
numerically accounting for shifts due to propagation delay and other
components in the PDH discriminator (Fig.$\:$\ref{fig:diode-mod_Bode}(b)).
The phase shift at $f_{\mathrm{UG}}=1.06\text{ MHz}$ is $-51.9^{\circ}$
and dominates our accounting of the overall loop phase delay (Table$\,$\ref{tab:phase}).
In addition to the possible phase lag from the laser diode driver
circuit \citep{Preuschoff22rosi}, we suspect the large lag at $1\text{ MHz}$
could be due to sub-optimal alignment of the optical feedback to the
diode \citep{Hall99MP} or laser diode thermal effects \citep{Hall99MP,Fox03NIST}.
Reference \citep{Oswald22Ths} offers an excellent discussion on the
phase lag introduced by laser diodes.

Note that the PDH error signal transfer function is a 1st-order low-pass
(Eq.$\:$\ref{eq:cavity-transfer-function}). However, the cavity
transmission response to laser frequency modulation is a 2nd-order
low-pass \citep{Houssin90rosi,Lawrence99josabj,Oswald22Ths}, making
it slower and more challenging to analyze.

\subsection{PDH demodulation electronics}

\label{appendix:lockin-transfer-function}

\begin{figure}
\includegraphics[width=1\columnwidth]{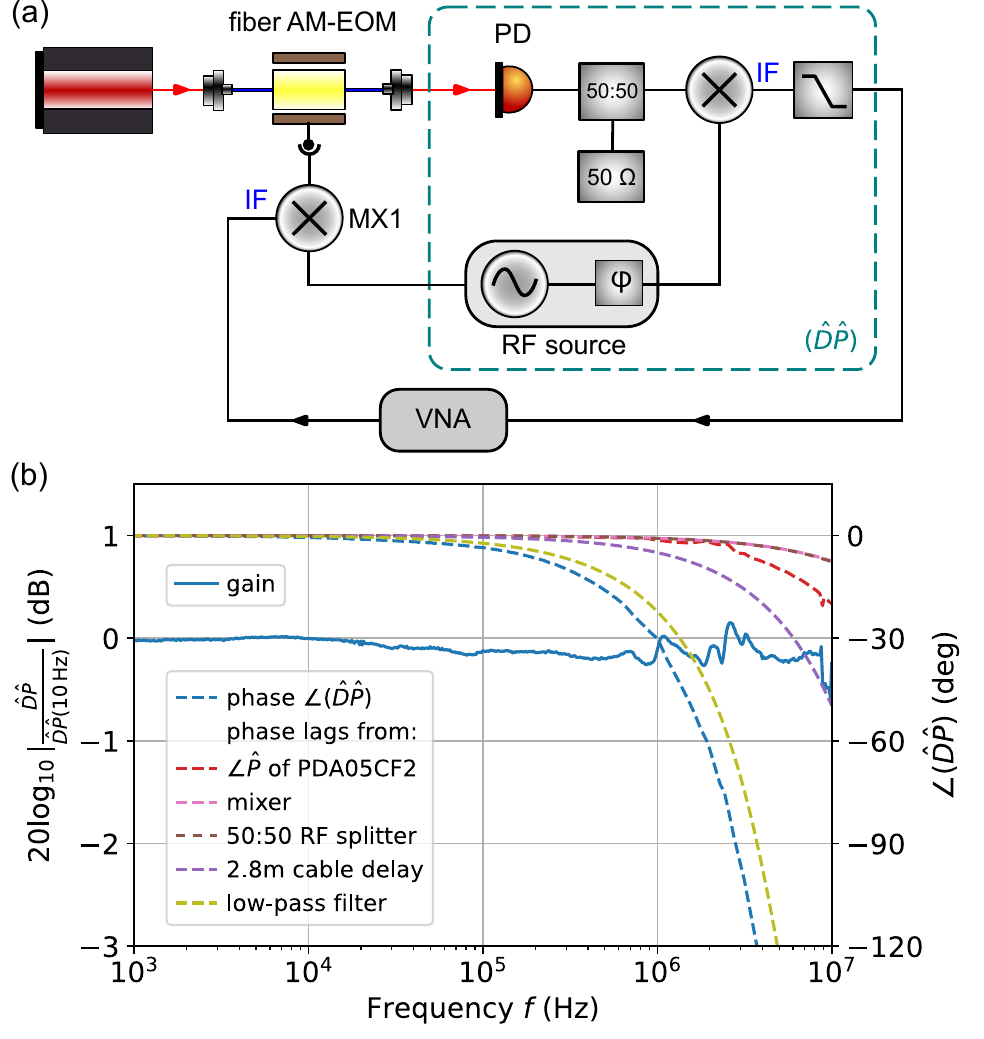}

\caption{\protect\label{fig:OFD_Bode}(a) The setup to measure the transfer
function $\hat{D}\hat{P}$ (components in the cyan dash box). AM-EOM
is the wideband EOM amplitude modulator. An extra mixer MX1 up-converts
VNA modulation to $\Omega/2\pi$ frequency. The transfer function
of MX1 is measured separately to compensate the measurement in (a).
(b) Bode plots of $\hat{D}\hat{P}$ (blue solid and dash lines) and
the phase lags from the components in $\hat{D}\hat{P}$ (dash lines
in other colors). Measurements of individual components are elaborated
in Appendix$\:$\ref{appendix:lockin-transfer-function}.}
\end{figure}

In this appendix we measure the transfer function of the PDH demodulation
electronics $\hat{D}\hat{P}$ consisting of PD $\hat{P}$ and mixer
plus low-pass $\hat{D}$. Recall the contribution of the quantity
$\hat{D}\hat{P}$ to the overall PDH discriminator transfer function
in Eq.$\:$\ref{eq:PDH-transfer}. The device under test is the enclosed
by the dashed cyan line in Fig.$\:$\ref{fig:OFD_Bode}(a).

Modulation in laser light is applied by the fast AM-EOM. An extra
mixer MX1 up-converts VNA modulation to $\Omega/2\pi$. We independently
check mixer transfer functions by cascading two identical mixers (one
up-convert, the other down-convert). After compensating the lag of
MX1 and propagation delay in fiber, we get $\hat{D}\hat{P}$. In Fig.$\:$\ref{fig:OFD_Bode}(b),
the phase lag traces of 50:50 RF splitter, low-pass filter and cable
delay are measured individually by VNA. $\angle\hat{L}_{\mathrm{PD}}$
of PD (PDA05CF2) is measured by VNA modulating the AM-EOM and then
$\angle\hat{P}$ is calculated from Eq.$\:$\ref{eq:PD-phase}. At
$f_{\mathrm{UG}}=1.06\text{ MHz}$ of the configuration 3, the sum
of individual lags are $-31.9^{\circ}$, which matches well with $\angle(\hat{D}\hat{P})=-32.8^{\circ}$.
This check further confirms that our transfer function model in Appendix
\ref{appendix:lock-in-detection} works and there is no unknown phase
lag in the lock-in detection. 

\section*{Bibliography}

\bibliographystyle{apsrev4-2}
\nocite{*}
\bibliography{pdh_Bode}

\begin{thebibliography}{83}
\bibitem [{no-()}]{no-endorse}%
  \BibitemOpen
  \href@noop {} {\bibinfo {title} {Products or companies named here are
  included in the interest of completeness and does not imply endorsement by
  the authors or by the {US Government}.}}\BibitemShut {Stop}%
\end{thebibliography}

\end{document}